\documentclass{aastex631}

\usepackage[utf8]{inputenc}
\usepackage{amsthm,amsmath,amssymb}
\usepackage{lipsum}
\usepackage{float}
\usepackage{soul}
\usepackage[T1]{fontenc}

\hypersetup{linkcolor=red,citecolor=blue,filecolor=cyan,urlcolor=magenta}

\begin{document}

\title{BSN-VI: Multiband Light Curve Modeling of Four W UMa-Type Contact Binaries\\
I. Revisiting Energy Transfer Mechanisms and Luminosity Behavior}

\author[0009-0006-1033-5885]{Elham Sarvari}
\altaffiliation{First author: elham.sarvari@aei.mpg.de}
\affiliation{BSN Project; Independent researcher, 12101 Berlin, Germany}

\author[0000-0002-0196-9732]{Atila Poro}
\altaffiliation{Corresponding author: atilaporo@bsnp.info, atila.poro@obspm.fr}
\affiliation{LUX, Observatoire de Paris, CNRS, PSL, 61 Avenue de l'Observatoire, 75014 Paris, France}
\affiliation{Department of Astronomy, Raderon AI Lab, Burnaby, British Columbia V5C 0J3, Canada}

\author[0000-0003-1263-808X]{Raul Michel}
\affiliation{Instituto de Astronom\'ia, UNAM. A.P. 106, 22800 Ensenada, BC, M\'exico}

\author[0000-0001-7069-7403]{Anna Francesca Pala}
\affiliation{European Southern Observatory, Karl Schwarzschild Straße 2, D-85748 Garching, Germany}

\author[0000-0002-3263-9680]{Mehmet Tanriver}
\affiliation{Department of Astronomy and Space Science, Faculty of Science, Erciyes University, Kayseri TR-38039, Türkiye}
\affiliation{Erciyes University, Astronomy and Space Science Observatory Application and Research Center, Kayseri TR-38039, Türkiye}

\author[0000-0002-7215-926X]{Ahmet Bulut}
\affil{Department of Physics, Faculty of Arts and Sciences, Çanakkale Onsekiz Mart University, Terzioğlu Kampüsü, TR-17020, Çanakkale, Türkiye}
\affil{Astrophysics Research Center and Observatory, Çanakkale Onsekiz Mart University, Terzioğlu Kampüsü, TR-17020, Çanakkale, Türkiye}

\author[0000-0002-9314-0648]{Ahmet Keskin}
\affiliation{Department of Astronomy and Space Science, Faculty of Science, Erciyes University, Kayseri TR-38039, Türkiye}

\author[0000-0003-0524-2204]{Mark G. Blackford}
\affiliation{Variable Stars South (VSS), Congarinni Observatory, Congarinni, NSW, 2447, Australia}

\begin{abstract}
We presented the first high-precision, detailed photometric analysis of four W Ursae Majoris (W UMa)-type contact binaries, Linear 10772300, Linear 11150338, Linear 20372537 and DM Cir. In addition to ground-based multiband photometric observations, data from the Transiting Exoplanet Survey Satellite (TESS) were employed for the analysis of the DM Cir system. New ephemeris and linear fit to the O–C diagrams were derived using extracted times of minima and additional literature. The light curve modeling was performed using the PHysics Of Eclipsing BinariEs (PHOEBE) Python code and the BSN application, employing a Markov Chain Monte Carlo approach. In each systems, the two stellar components exhibited minimal temperature differences ($\Delta T<150$ K), confirming efficient energy exchange within their common convective envelopes. Absolute parameters were estimated using the Gaia Data Release 3 (Gaia DR3) parallax and astrophysical equations. Based on effective temperatures and component masses, two systems were classified as W-subtype systems, while others belonged to the A-subtype. We computed the initial masses of the primary ($M_{1i}$) and secondary ($M_{2i}$) components for four target systems using a method based on the observational properties of overluminous secondary components. We found initial primary masses in the range 0.6–1.0$M_\odot$ and initial secondary masses in the range 0.9–1.7$M_\odot$ with mass loss $<1.0M_{\odot}$. We investigated the relative energy transfer rates ($U_{1}$ and $U_{2}$) and nuclear luminosities ($L_{10}$ and $L_{20}$) based on the physical parameters of 411 W UMa–type contact binaries, including the four systems analyzed in this study, through wide range of mass ratios. The results for all systems provided a comprehensive view of energy transfer behavior throughout different evolutionary stages of contact binaries.
\end{abstract}


\section{Introduction}
\label{sec1}
Binary stars play a fundamental role in astrophysics, not only because they are numerous, but also because they serve as a primary source of our knowledge about the fundamental properties of stars (\citealt{kallrath2009eclipsing, latham1992spectroscopic}). Observational studies indicate that more than half of the stars in the solar neighborhood belong to binary or multiple systems (\citealt{kallrath2009eclipsing}). Binary systems are characterized by measurable gravitational interactions that allow stellar parameters to be estimated. Orbital motion produces detectable variations in position and velocity over a wide range of stellar separations and luminosity ratios, allowing both stars to be studied in detail depending on their distances, brightnesses, and motions (\citealt{latham2002survey}).

Eclipsing binary systems are especially valuable in this context, as their orbital planes are oriented edge-on to the observer, allowing eclipses to occur. W UMa systems are a subclass of eclipsing binaries in which both stars fill their inner Lagrangian surfaces (usually referred to as Roche Lobes), and share a common convective envelope (\citealt{lucy1968structure}). If the components fill their Roche lobes, the system is classified as a contact binary, and its degree of filling can be quantified by the filling factor (\citealt{wilson2001binary}). W UMa systems are commonly divided into two subclasses: A and W subtypes. In A-subtype systems, the more massive star is hotter; whereas in W-subtype systems, the more massive star is cooler than its companion (\citealt{binnendijk1970orbital,poro2026-27systems}). Both subclasses are characterized by continuously varying light curves, with only a small difference between the depths of the minima. The nearly equal depths of the two minima indicate that, despite their different masses, both components have nearly identical temperatures (\citealt{qian2014optical}). However, some systems exhibit notable asymmetries—such as the O'Connell effect \cite{o1951so}, where the two maxima differ in brightness. Such asymmetries are more frequently observed in W-subtype systems (\citealt{kallrath2009eclipsing}). 

Despite extensive studies, W~UMa-type contact binary systems remain under active investigation, and precise observations and analysis are crucial for understanding their structure and evolution (\citealt{rahunen1982origin}). Establishing a consistent theoretical framework for these systems is challenging, as their evolution is governed by the coupled effects of mass and energy transfer, angular momentum loss (AML), and the requirement to maintain an apparent thermal equilibrium at the stellar surfaces (\citealt{li2004structure,stepien2005evolutionary}). These processes operate on different timescales and are not directly constrained by observations, leading to degeneracies among model parameters and complicating the interpretation of observed systems. Moreover, the evolutionary states of the components are often assessed or compared using single-star evolutionary models, despite the fact that ongoing interaction between the components can significantly alter their internal structure and evolutionary paths (\citealt{gazeas2008angular}).

This study presents ground-based, multiband photometric observations of four W UMa-type contact binaries, with the aim of conducting a detailed photometric analysis. It continues the investigation initiated by \cite{2025b}, \cite{poro2025bsn}, and \cite{2025c} by conducting new observations and a comprehensive analysis of W UMa-type contact binary systems in the BSN project\footnote{\url{https://bsnp.info}}. The structure of this paper is organized as follows: Section \ref{sec2} provides details about the target systems. Section \ref{sec3} describes the ground-based observations and the data reduction process. Section \ref{sec4} presents the determination of a new ephemeris, while Section \ref{sec5} includes the photometric light curve analysis for the target systems. Section \ref{sec6} outlines the methods used to determine the absolute parameters. Finally, Section \ref{sec7} offers a discussion and conclusions.

\vspace{0.6cm}
\section{Target Systems}
\label{sec2}
We investigated four binary star systems: Linear 10772300 (hereinafter L10772300), Linear 11150338 (hereinafter L11150338), Linear 20372537 (hereinafter L20372537), and DM Cir. These systems are classified as contact binaries in both the All-Sky Automated Survey for SuperNovae (ASAS-SN; \citealt{shappee2014man}; \citealt{jayasinghe2018asas}) and the AAVSO Variable Star Index (VSX\footnote{\url{https://vsx.aavso.org/}}) catalogs. Table \ref{targets} presents some specifications of the target systems as listed in the Gaia DR3\footnote{\url{https://www.cosmos.esa.int/web/gaia/data-release-3}} catalog. Table \ref{observations} provides the observation characteristics for each system, including the observation date, filters, exposure time, and the name of the observatory. In each case, we utilized $V_{\text{max}}$ derived from our observations. Brief introductions to the targets are provided below.

$\bullet$ L10772300: The binary system L10772300 was discovered by the Lincoln Near-Earth Asteroid Research (LINEAR) project and included in the LINEAR III Catalog of Periodic Variables (\citealt{palaversa2013exploring}). 
In addition to ASAS-SN and VSX, it is also classified as a contact binary in several other catalogs, including the Catalina Surveys catalog of periodic variable stars (CRTS; \citealt{drake2014catalina}), the TESS Input Catalog version 8.2 (TIC; \citealt{paegert2021tess}), the Sloan Digital Sky Survey (SDSS; \citealt{ahumada202016th}), and the Zwicky Transient Facility (ZTF; \citealt{bellm2019zwicky}). The orbital period is 0.244452 days, while the effective temperature is reported as 5233 K in Gaia DR3 and $5138\pm153$ K in the TIC.

$\bullet$ L11150338: It is part of the LINEAR III survey's sample of variable stars spanning 10,000 $deg^2$ of the northern sky. An orbital period of 0.314730 days has been reported by both the VSX catalog and the catalog of contact binaries in LAMOST DR7 (\citealt{qian2020contact}), while the time of minimum is available from the ASAS-SN catalog. The effective temperature of the system is reported as 6048 K by Gaia DR3 and $5980\pm87$ K by the TIC.

$\bullet$ L20372537: The ZTF Variable Stars Catalog and the VSX catalog report orbital periods of 0.2887654 and 0.288760 days, respectively. 
The TIC catalog reports an effective temperature of $4661\pm154$ K for this system.

$\bullet$ DM Cir: Houk, N., \& Cowley, A. P. (\citeyear{houk1975university}) reported this system as a new variable from the Southern hemisphere for the first time. The orbital period of DM Cir is reported as 0.386777 days in the ASAS catalog of eclipsing binaries with RASS counterparts (\citealt{szczygiel2008coronal}), 0.3867749 days in the General Catalogue of Variable Stars (GCVS; \citealt{malkov2006catalogue}), 0.386773 days in the VSX catalog, and 0.3868075 days in the TESS. 

\begin{table*}
\renewcommand\arraystretch{1.2}
\centering
\caption{General parameters of the target systems, including coordinates, distances, and orbital periods.}
\begin{tabular}{cccccc}
\hline
System & TIC & RA$.^\circ$(J2000) & DEC$.^\circ$(J2000) & d(pc) & $P_d$ (day)\\
\hline
L10772300  & 23871740  & 211.7077137  &  40.7781799 & 1292(59)  &  0.244452(VSX)\\
L11150338  & 138186324  & 212.1970867  & 35.3223105  & 2151(126)  &  0.314730(VSX)\\
L20372537  & 101717663  & 259.8652614  &  36.2387627 & 1252(139)  &  0.288766(VSX)\\
DM Cir  & 47309500  & 231.0349806  & -56.8376572 & 163(1)  &  0.386777(ASAS-SN)\\
\hline
\end{tabular}
\label{targets}
\end{table*}

\vspace{0.6cm}
\section{Observation and Data Reduction}
\label{sec3}
Photometric observations and data reduction procedures were conducted using standard filters for four target systems at two observatories located in the Northern and Southern Hemispheres. Three of the systems were observed at the San Pedro Mártir Observatory (SPM) in México, while the remaining system was observed at the Congarinni Observatory in Australia. Among the targets, DM Cir was the only system with available timeseries data from TESS (\citealt{ricker2010transiting}, \citealt{stassun2018tess}). The observational details, including dates, filters, and exposure times, are provided in Table \ref{observations}, while the characteristics of the comparison and check stars used during the observations are summarized in Table \ref{stars}. A detailed description of the observational facilities is presented in the following section.

\subsection{SPM Observatory}
The three binary systems L10772300, L11150338, and L20372537 were observed at the San Pedro Mártir (SPM) Observatory in México. This observatory is located at a longitude of {$115^\circ$ $27'$ $49''$} W, latitude {$31^\circ$ $02'$ $39''$} N, and an altitude of 2830 meters above sea level. Observations were conducted using a 1.5-meter telescope 
equipped with the RUCA filter wheel and a Spectral Instruments CCD camera. The detector consists of an e2v CCD42-40 chip with $13.5 \times 13.5~\mu\mathrm{m}^2$ pixels, a gain of $1.39e^{-}\mathrm{ADU}^{-1}$, and a readout noise of $3.54~e^{-}$. Standard Johnson–Cousins $B$, $V$, $R_c$, and $I_c$ filters were employed during the observations. Photometric image processing was performed using standard IRAF routines (\citealt{tody1986iraf}), and data reduction was carried out using bias subtraction and flat-field correction.

\vspace{0.3cm}
\subsection{Congarinni Observatory}
Photometric observations of DM Cir were conducted in June 2019 and May 2020 using standard $B$, $V$, and $I_c$ filters. These observations were performed with an Orion ED80T CF refractor telescope at the Congarinni Observatory in Australia, located at longitude {$152^\circ$} $52'$ E and latitude {$30^\circ$} $44'$ S. An Atik One 6.0 CCD camera was used with $1 \times 1$ binning. Exposure times were set to 30 seconds for the $B$ filter and 12 seconds for both the $V$ and $I_c$ filters. A total of 1935 images were acquired in the $BVI_c$ filters throughout the observational campaign.

\vspace{0.3cm}
\subsection{TESS Observations}
NASA launched the TESS Satellite in 2018 to discover TESS, equipped with four wide-field cameras, systematically observes different regions of the sky, dedicating 27.4 days to each sector. In this study, time-series photometric data from TESS were available only for the DM Cir system, as comparable data were not available for the other three targets. Observations from Sectors 38, 39, and 65, obtained in 2021 and 2023, were used. Sectors 38 and 39 were observed with an exposure time of 600 seconds, while Sector 65 used a shorter exposure time of 200 seconds. The times of minima extracted from these TESS sectors for DM Cir are presented in Table \ref{appendix-table1}. All data were retrieved from the Mikulski Archive for Space Telescopes (MAST)\footnote{\url{https://mast.stsci.edu/portal/Mashup/Clients/Mast/Portal.html}}, and the light curves were detrended using the SPOC pipeline.

\begin{table*}
\renewcommand\arraystretch{1.2}
\caption{Specifications of the ground-based observations.}
\centering
\begin{center}
\footnotesize
\begin{tabular}{c c c c c}
\hline
System & Observation(s) Date & Filter & Exposure time(s) & Observatory\\
\hline
L10772300 &	2024(May7) & $\mathit{BVR_cI_c}$ & $\mathit{B(60),V(30),R_c(20),I_c(20)}$ & SPM\\
L11150338 &	2024(May13) & $\mathit{BVR_cI_c}$ & $\mathit{B(30),V(15),R_c(10),I_c(10)}$ & SPM \\
L20372537 &	2024(June17) & $\mathit{BVR_cI_c}$  & $\mathit{B(60),V(30),R_c(20),I_c(15)}$ & SPM\\
DM Cir & 2019(June19,21,22), 2020(May10,11) & $\mathit{BVI_c}$ & $\mathit{B(30),V(12),I_c(12)}$ & Congarinni\\
\hline
\end{tabular}
\end{center}
\label{observations}
\end{table*}

\begin{table*}
\renewcommand\arraystretch{1.2}
\caption{List the comparisons and check stars in the ground-based observations. Coordinates come from the Gaia DR3.}
\centering
\begin{center}
\footnotesize
\begin{tabular}{c c c c c}
\hline
System & Star Type & RA$.^\circ$(J2000) & DEC$.^\circ$(J2000)\\
\hline
L10772300	&	Comparison	& 211.713538 & 40.808615 \\
L10772300	&	Check	& 211.695935 & 40.778505 \\
L11150338	&	Comparison	& 212.190099 & 35.291748 \\
L11150338	&	Check	& 212.246209 &	35.324273 \\
L20372537	&	Comparison	& 259.927774 & 36.257595 \\
L20372537	&	Check &	259.861978 & 36.255877 \\
DM Cir	&	Comparison	& 231.010255 & -56.676385	\\
DM Cir	&	Check	&	231.057119	& -56.699295	\\
\hline
\end{tabular}
\end{center}
\label{stars}
\end{table*}

\vspace{0.6cm}
\section{New Ephemeris}
\label{sec4}
The primary and secondary minima of all four systems were extracted from our ground-based observations with standard $B, V, R_c,$ and $I_c$ filters. They were also combined with 408 mid-eclipse times extracted from three time-series TESS sector data, complemented by additional minima collected from the literature for the subsequent calculations.
Since the times of minima were reported in both Barycentric Julian Date in Barycentric Dynamical Time $(BJD_{TDB})$ and Heliocentric Julian Date (HJD), we first converted all minima times to $BJD_{TDB}$ using an online converter\footnote{\url{https://astroutils.astronomy.osu.edu/time/hjd2bjd.html}}(\citealt{eastman2010achieving}). Gaussian curve fitting was applied to derive the precise times of minima. The extracted times of minima are listed in Table \ref{tab-min}. Additionally, mid-eclipse times from the TESS data in sectors 38, 39, and 65 (observed in 2021 and 2023) for DM Cir are listed in Appendix Table \ref{appendix-table1}.

\begin{figure*}
\centering
\includegraphics[width=0.45\textwidth]{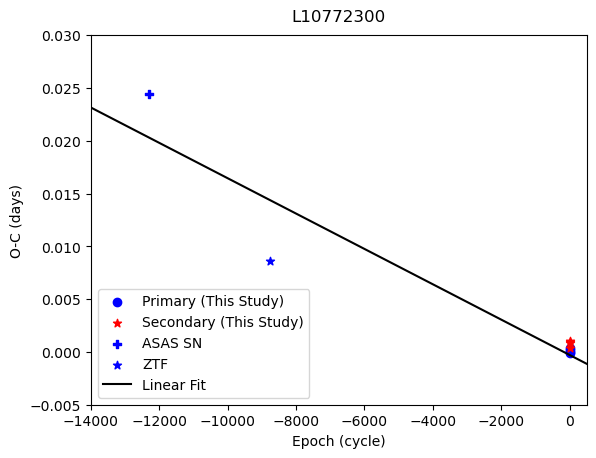}
\includegraphics[width=0.45\textwidth]{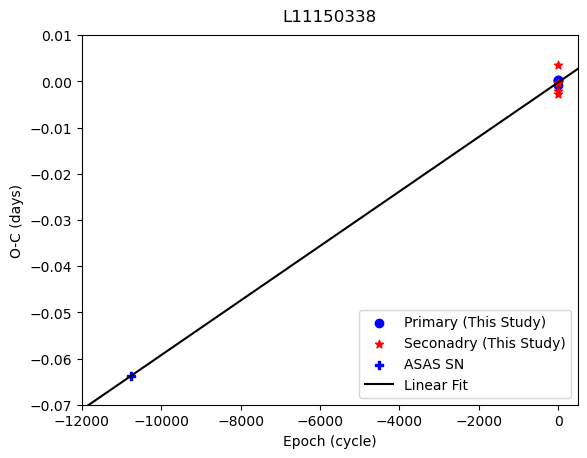}
\includegraphics[width=0.45\textwidth]{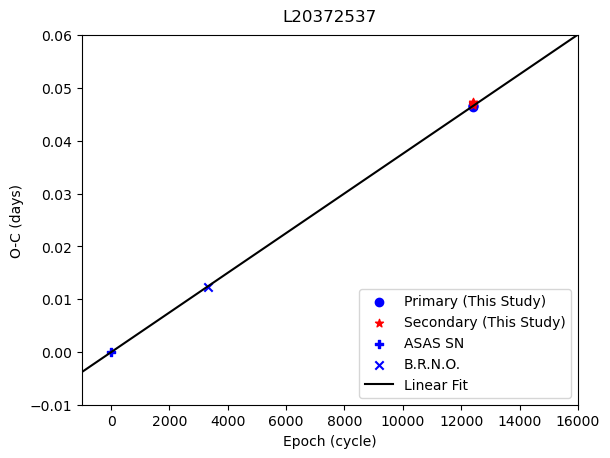}
\includegraphics[width=0.45\textwidth]{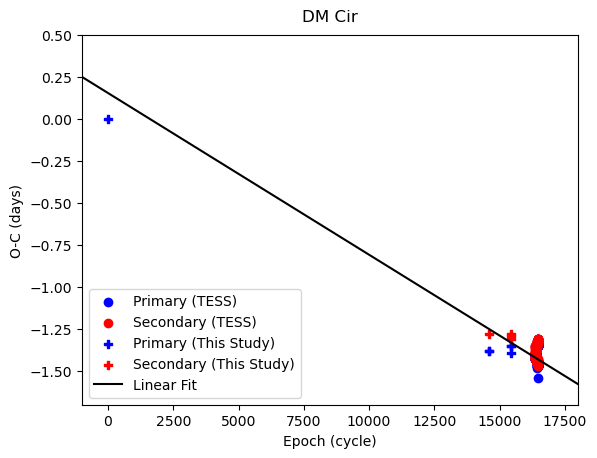}
\caption{O–C diagrams illustrating the period variations of the targets.}
\label{fig-O-C}
\end{figure*}

In binary star systems, the Observed–Calculated (O-C) diagram is an essential tool for refining and updating the system's ephemeris. In this study, a linear fit is more appropriate for the O-C diagrams, given the limited number of observations and minima available for our four binary systems. The O-C diagrams are presented in Figure \ref{fig-O-C}. For each system, the epoch and O-C values were calculated using the ephemeris references listed in Table \ref{tab-ephemeris}.
The O-C values were calculated based on the linear ephemeris

\begin{equation}
\mathrm{BJD}(E) = \mathrm{BJD}_{0} + P \times E,
\end{equation}
where $\mathrm{BJD}(E)$ denotes the predicted eclipse time at epoch $E$, $\mathrm{BJD}_{0}$ is the reference epoch, and $P$ represents the orbital period. For L10772300 and L11150338, we used minima from our ground-based observations and orbital periods reported in the VSX catalog. For L20372537, the reference ephemeris was determined using a minimum time from \cite{jayasinghe2019asas} and a period from the ASAS-SN catalog. For DM Cir, both the minimum and period were taken from \cite{otero2004new}. The new ephemeris for each system is presented in Table \ref{tab-ephemeris}.

\begin{table*}
\renewcommand\arraystretch{1.2}
\caption{The times of minima for four target systems from both the literature and this study}
\centering
\small
\begin{tabular}{c c c c c c c}
\hline
System & Min.($BJD_{TDB}$) & Error & Filter & Epoch & O-C & Reference \\
\hline
L10772300 & 2457433.02001 &  &  & -12296 & 0.0244 & ASAS-SN\\
& 2458295.18642 &  &  & -8769 & 0.0086 & ZTF \\
& 2460438.77773 & 0.00019 & $\textit{B}$ & 0 & 0.0003 & This study\\
& 2460438.77739 & 0.00016 & $\textit{I}$ & 0 & 0 & This study \\
& 2460438.77761 & 0.00016 & $\textit{R}$ & 0 & 0.0002 & This study \\
& 2460438.77732 & 0.00013 & $\textit{V}$ & 0 & -0.0001 & This study \\
& 2460438.89997 & 0.00031 & $\textit{B}$ & 0.5 & 0.0004 & This study \\
& 2460438.90058 & 0.00018 & $\textit{I}$ & 0.5 & 0.0010 & This study \\
& 2460438.90063 & 0.00014 & $\textit{R}$ & 0.5 & 0.0010 & This study \\
& 2460438.90048 & 0.00010 & $\textit{V}$ & 0.5 & 0.0009 & This study \\
    
L11150338  & 2457061.053915 &  &  & -10751 & -0.0637 & ASAS-SN\\
& 2460444.78019 & 0.00029 & $\textit{B}$ & 0 & 0.0004 & This study \\
& 2460444.77897 & 0.00046 & $\textit{I}$ & 0 & -0.0009 & This study \\
& 2460444.77983 & 0.00031 & $\textit{R}$ & 0 & 0 & This study \\
& 2460444.78014 & 0.00023 & $\textit{V}$ & 0 & 0.0003 & This study \\
& 2460444.94065 & 0.00120 & $\textit{B}$ & 0.5 & 0.0035 & This study \\
& 2460444.93435 & 0.00227 & $\textit{I}$ & 0.5 & -0.0028 & This study \\
& 2460444.93520 & 0.00149 & $\textit{R}$ & 0.5 & -0.0020 & This study \\
& 2460444.93684 & 0.00141 & $\textit{V}$ & 0.5 & -0.0004 & This study \\

L20372537  & 2456900.79026 &  &  & 0 & 0 & ASAS-SN\\
& 2457856.59818 & 0.00001 &  & 3310 & 0.0123 & VarAstro\\
& 2460479.72839 & 0.00015 & $\textit{B}$ & 12394 & 0.0467 & This study\\
& 2460479.72801 & 0.00009 & $\textit{I}$ & 12394 & 0.0463 & This study\\
& 2460479.72832 & 0.00011 & $\textit{R}$ & 12394 & 0.0466 & This study\\
& 2460479.72826 & 0.00008 & $\textit{V}$ & 12394 & 0.0466 & This study\\
& 2460479.87293 & 0.00019 & $\textit{B}$ & 12394.5 & 0.0469 & This study\\
& 2460479.87320 & 0.00013 & $\textit{I}$ & 12394.5 & 0.0471 & This study\\
& 2460479.87329 & 0.00034 & $\textit{R}$ & 12394.5 & 0.0472 & This study\\
& 2460479.87332 & 0.00014 & $\textit{V}$ & 12394.5 & 0.0472 & This study\\

DM Cir  & 2458654.13468  & 0.00024 & $\textit{V}$ & 14588  & -0.0199 & This study\\
& 2458656.06857 & 0.00018 & $\textit{V}$ & 14593  &  -0.0199 & This study \\
& 2458657.03693 & 0.00036 & $\textit{V}$ & 14595.5 & -0.0185 & This study \\
& 2458980.96116 & 0.00024 & $\textit{V}$ & 15433  &  -0.0200 & This study \\
& 2458980.96169 & 0.00026 & $\textit{B}$ & 15433  & -0.0195 & This study \\
& 2458980.96183 &  0.00025 & $\textit{I}$ & 15433  & -0.0194 & This study \\
& 2458981.15578 &  0.00025 & $\textit{V}$ & 15433.5 & -0.0188 & This study\\
& 2458981.15602 &  0.00021 & $\textit{B}$ & 15433.5 & -0.0186 & This study\\
& 2458981.15618 & 0.00030 & $\textit{I}$ & 15433.5  & -0.0184 & This study\\
\hline
\label{tab-min}
\end{tabular}
\end{table*}

\begin{table*}
\renewcommand\arraystretch{1.2}
\caption{Reference and updated ephemerides of the four target systems.}
\centering
\begin{center}
\begin{tabular}{c c c}
\hline
System & Reference Ephemeris & updated Ephemeris\\
\hline
L10772300 & $2460438.77739 + 0.244452\times E$  & $2460438.77710(177) + 0.2444503(3) \times E$ \\
L11150338 & $2460444.77983 + 0.314730\times E$  & $2460444.77978(28) + 0.31473592(6) \times E$ \\
L20372537 & $2456900.79026 + 0.288760\times E$  & $2456900.79021(12) + 0.28876376(1) \times E$ \\
DM Cir & $2453011.85175 + 0.386777\times E$  & $2453011.85946(16) + 0.38677526(1) \times E$ \\
\hline
\end{tabular}
\end{center}
\label{tab-ephemeris}
\end{table*}

\vspace{0.6cm}
\section{Light Curve Solutions}
\label{sec5}
The light curve solution process was initiated by converting the time into orbital phase, using the new ephemeris listed in Table \ref{tab-ephemeris}. The analysis of contact binary systems was then conducted using PHOEBE version 2.4.9, a Python-based modeling tool \cite{prvsa2005computational, prvsa2016physics, conroy2020physics}, together with the BSN application version 1.0 \cite{paki2025bsn} offered by the BSN project, and Markov chain Monte Carlo (MCMC) approach (\citealt{prvsa2016physics}). This work provides the first photometric light curve solutions of the four target systems, which have not previously been studied in the literature. The contact binary systems approach was utilized based on catalog classifications and further supported by the short orbital periods along with the morphological features of the light curves.
The initial modeling of each target system was conducted using the PHOEBE code, which employs built-in optimization algorithms to estimate parameters before applying the MCMC approach. In all models, the bolometric albedos and gravity-darkening coefficients were set to $A_1=A_2=0.5$ (\citealt{rucinski1969proximity}) and $g_1=g_2=0.32$ (\citealt{lucy1967gravity}), respectively. The stellar atmospheres were modeled based on \cite{castelli2004new} study, while the limb darkening coefficients were treated as free parameters.

The initial components’ effective temperatures for three binary systems, L10772300, L11150338, and DM Cir, were obtained from Gaia DR3. As Gaia DR3 does not provide this information for L20372537, its temperature was instead obtained from the MAST archive with source information from TIC version 8.2. In all cases, the initial temperature was placed in the hotter component, based on the depths of the light curve minima. The effective temperature of the cooler component was estimated using the difference in depth between the primary and secondary minima of the light curves.

\begin{figure*}
\centering
\includegraphics[width=0.9\textwidth]{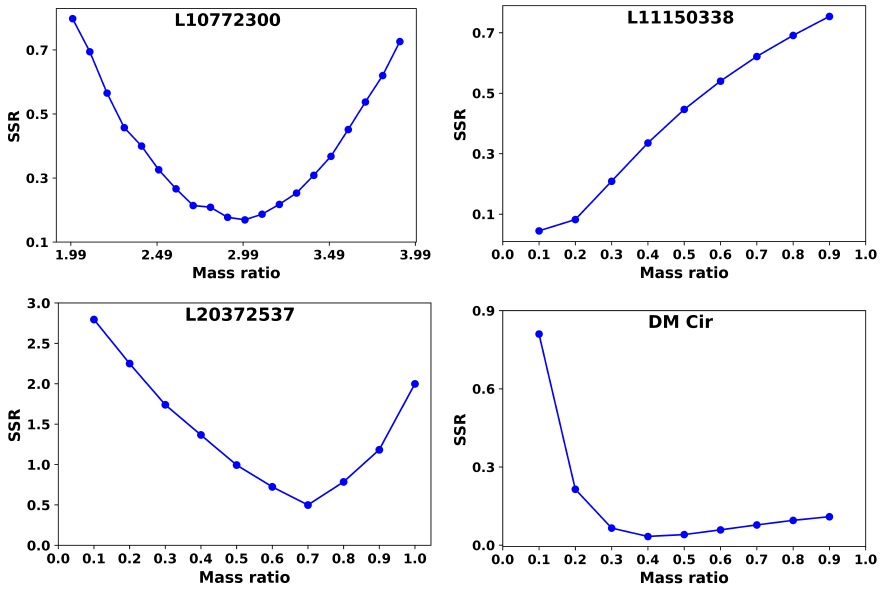}
\caption{Sum of the squared residuals as a function of the mass ratio.}
\label{fig-q}
\end{figure*}

The mass ratios $q = M_2/M_1$ of the systems were determined using the $q$-search method, based on photometric observations and following standard procedures (\citealt{terrell2005photometric}).
An initial range of $q=0.1$ to $q=20$ was explored for all targets. Based on the minima in the sum of squared residuals, the interval was subsequently narrowed for a more refined analysis. As shown in Figure \ref{fig-q}, each $q$-search curve exhibits a well-defined minimum, indicating the optimal mass ratio.

Our observations reveal an unequal brightness between the primary and secondary maxima in L10772300 and DM Cir target systems. Both cool and hot starspot configurations were examined on the secondary component in order to reproduce the observed light curve asymmetry. A cool starspot was ultimately adopted in the final light curve modeling, as it provided a better fit according to the reduced $\chi^{2}$ statistic. The most probable explanation for this asymmetry in the light curve's maxima is that the magnetic activity of the components is causing the presence of the starspot(s), which are introduced with the O'Connell effect (\citealt{o1951so}). Other explanations for the O’Connell effect have also been proposed (\citealt[e.g.,][]{1990ApJ...355..271Z,2003ChJAA...3..142L,2025ApJ...994....7F}), addressing different possible mechanisms such as mass transfer. Starspots are generally categorized as cold or hot based on their relative temperature to the stellar photosphere. In light curve modeling, whether to use a cold- or a hot-starspot depends on which process (photospheric magnetic fields or mass exchange between components) most consistently explains the physical causes of the O’Connell effect (\citealt{knote2022characteristics, kouzuma2019starspots}). Cool spots, often found on late-type secondary components, are generally attributed to photospheric magnetic fields. According to dynamo theory \cite{moss2004dynamo}, magnetic energy is generated by the conversion of kinetic energy that comes from convection in the outer layer. On the other hand, if a hot spot is considered in the modeling, it results from the impact of mass transfer between the components rather than from magnetic activity (\citealt{park2013light}). This scenario is more likely in contact and semi-detached systems, where at least one star fills its Roche lobe (\citealt{kouzuma2019starspots}). Colatitude ($Col.^{\circ}$), longitude ($Long.^{\circ}$), angular radius ($Radius.^{\circ}$), and the ratio of temperature ($T_{spot}/T_{star}$) are the characteristics that are identified for a starspot (Table \ref{tab-solutions}).

Next, we aimed to determine an acceptable theoretical fit for the observational data by utilizing the initial values. Additionally, we employed PHOEBE's optimization tools to refine the light curve solutions, using the same configuration described earlier. This allowed us to generate the results needed for the subsequent analysis step.
We applied the high-speed MCMC algorithm implemented in the BSN application to refine the light curve solutions. This approach allowed us to derive the final results along with their associated uncertainties.

The BSN application is well-suited for this analysis, as its procedures are consistent with the PHOEBE standard program and produce numerically agreement results (\citealt{paki2025bsn}). Both the PHOEBE code and BSN utilize the emcee package \cite{goodman2010ensemble} for MCMC computations; however, the BSN application significantly improves MCMC performance. Thanks to its implementation in modern .NET technologies, this application generates synthetic light curves over 40 times faster than PHOEBE. This performance advantage is maintained despite methodological differences. Both BSN and PHOEBE discretize the Roche geometry using surface meshes; however, BSN follows the formulation of \cite{mochnacki1984accurate}, while PHOEBE adopts a Wilson–Devinney–based surface parametrization scheme (\citealt{wilson1971realization}). The MCMC process in the BSN application operates based on five main parameters, $T_1$, $T_2$, $q$, $f$, and $i$. We employed 24 walkers and 1500 iterations to sample these parameters effectively. The third light component ($l_3$), representing a possible luminous contribution from the third component to the system's total flux, was found to be insignificant. For each of the four systems analyzed, the best theoretical fit to the observational data was obtained without including this additional component, indicating no detectable evidence of a third star contributing to the system's luminosity.

\begin{figure*}
\centering
\includegraphics[width=0.9\textwidth]{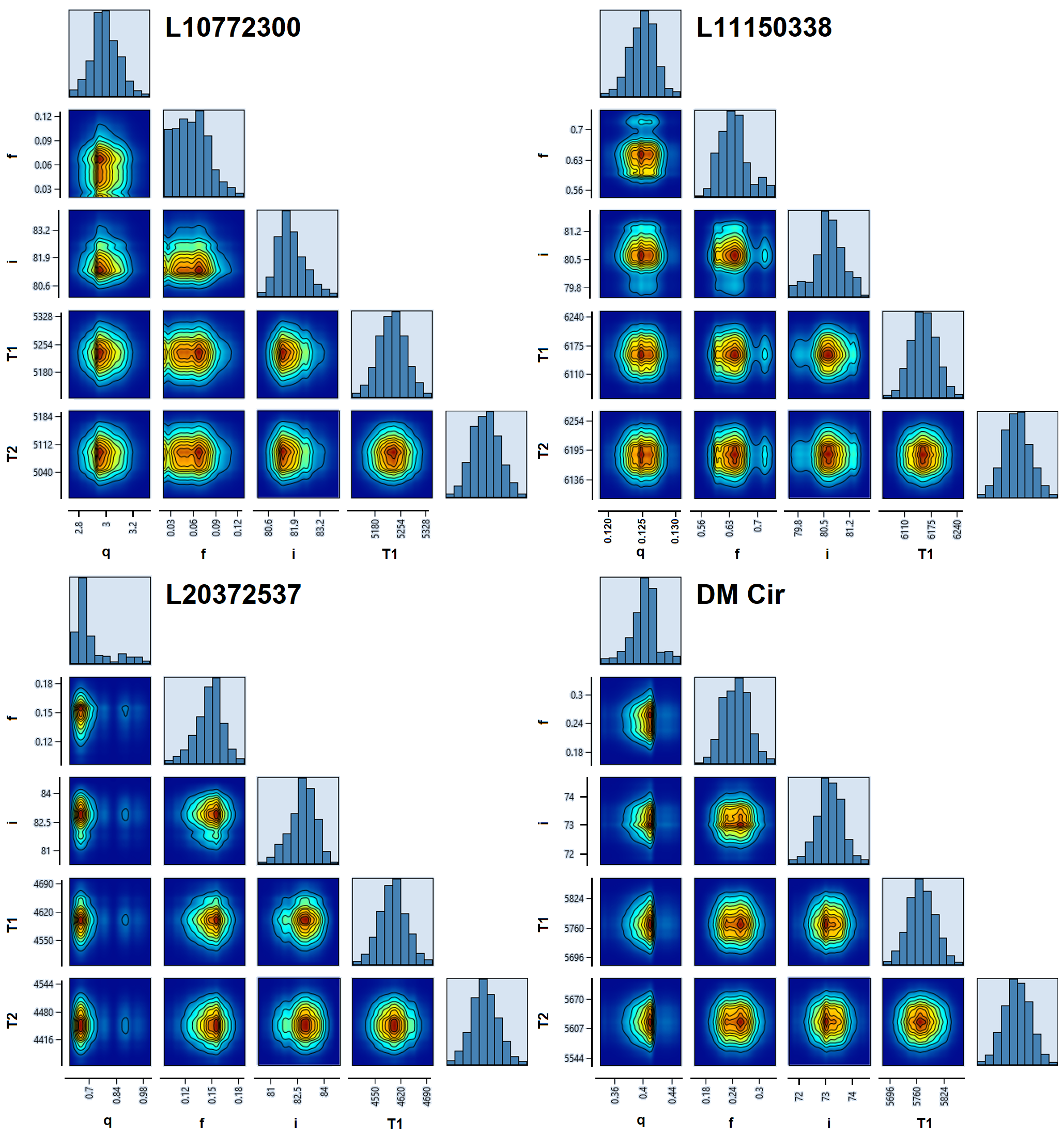}
\caption{Corner plots based on the heat-map of the target contact binary systems were determined through MCMC modeling by the BSN application.}
\label{fig-MCMC}
\end{figure*}

\begin{figure*}
\centering
\includegraphics[width=0.54\textwidth]{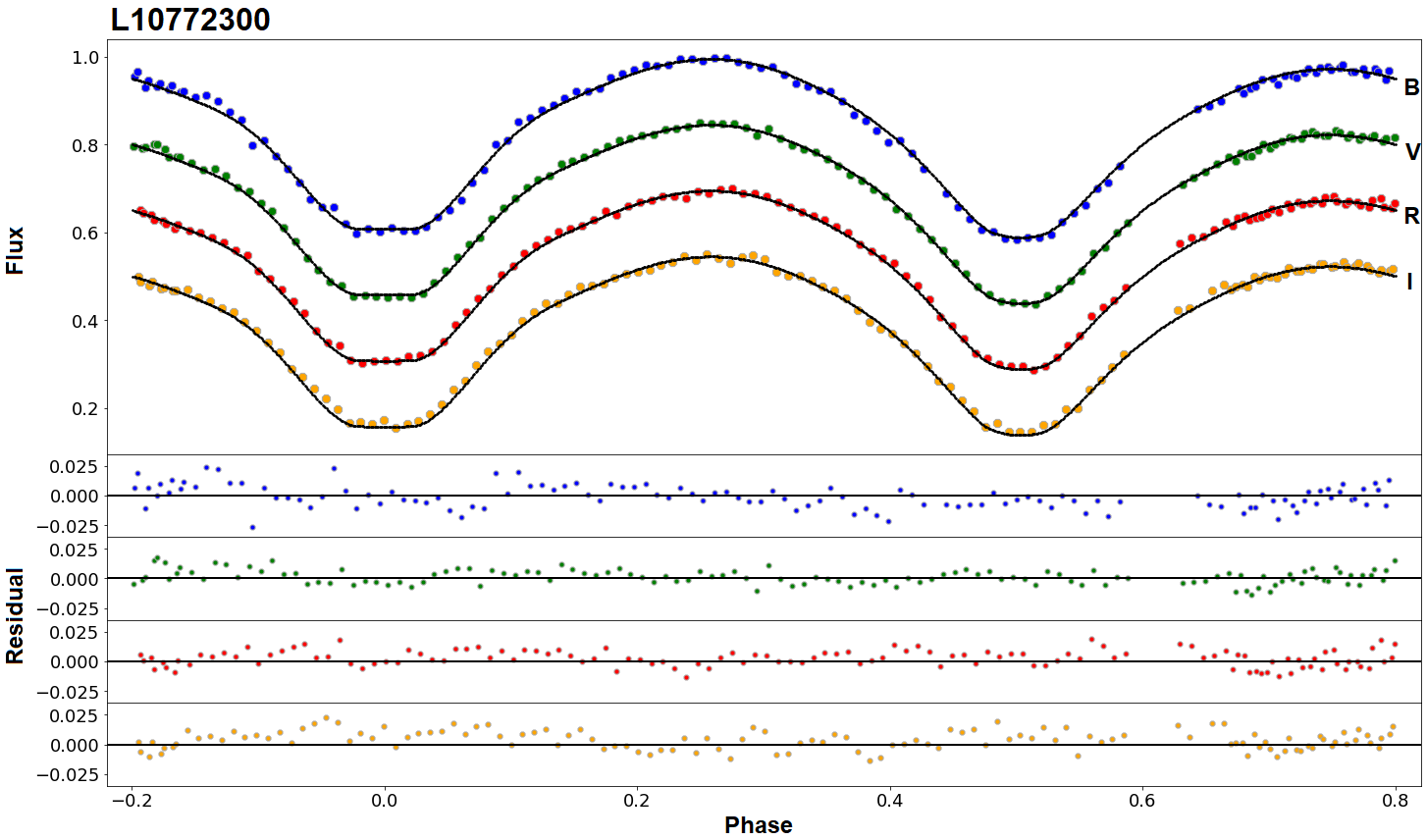}
\includegraphics[width=0.54\textwidth]{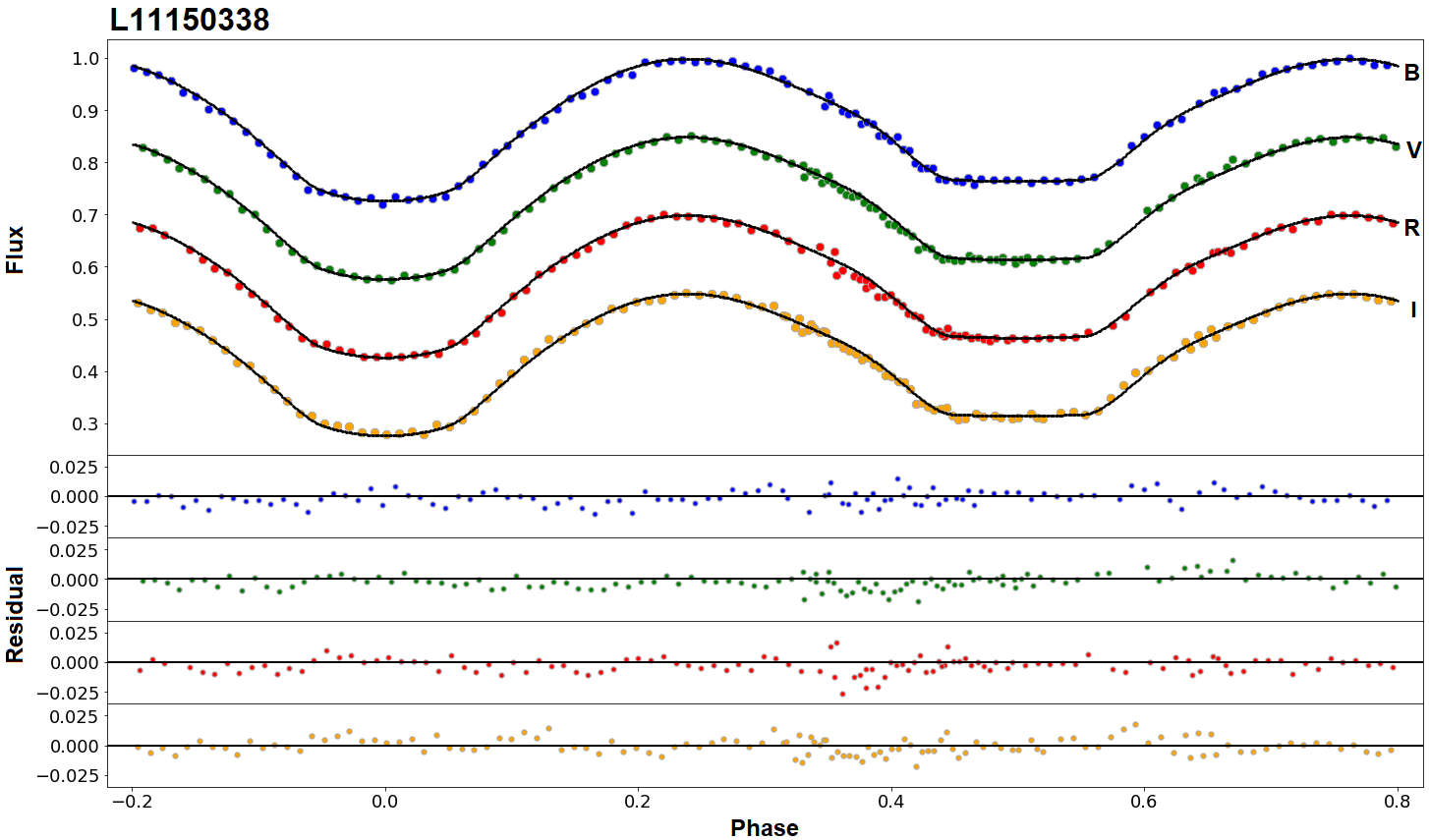}
\includegraphics[width=0.54\textwidth]{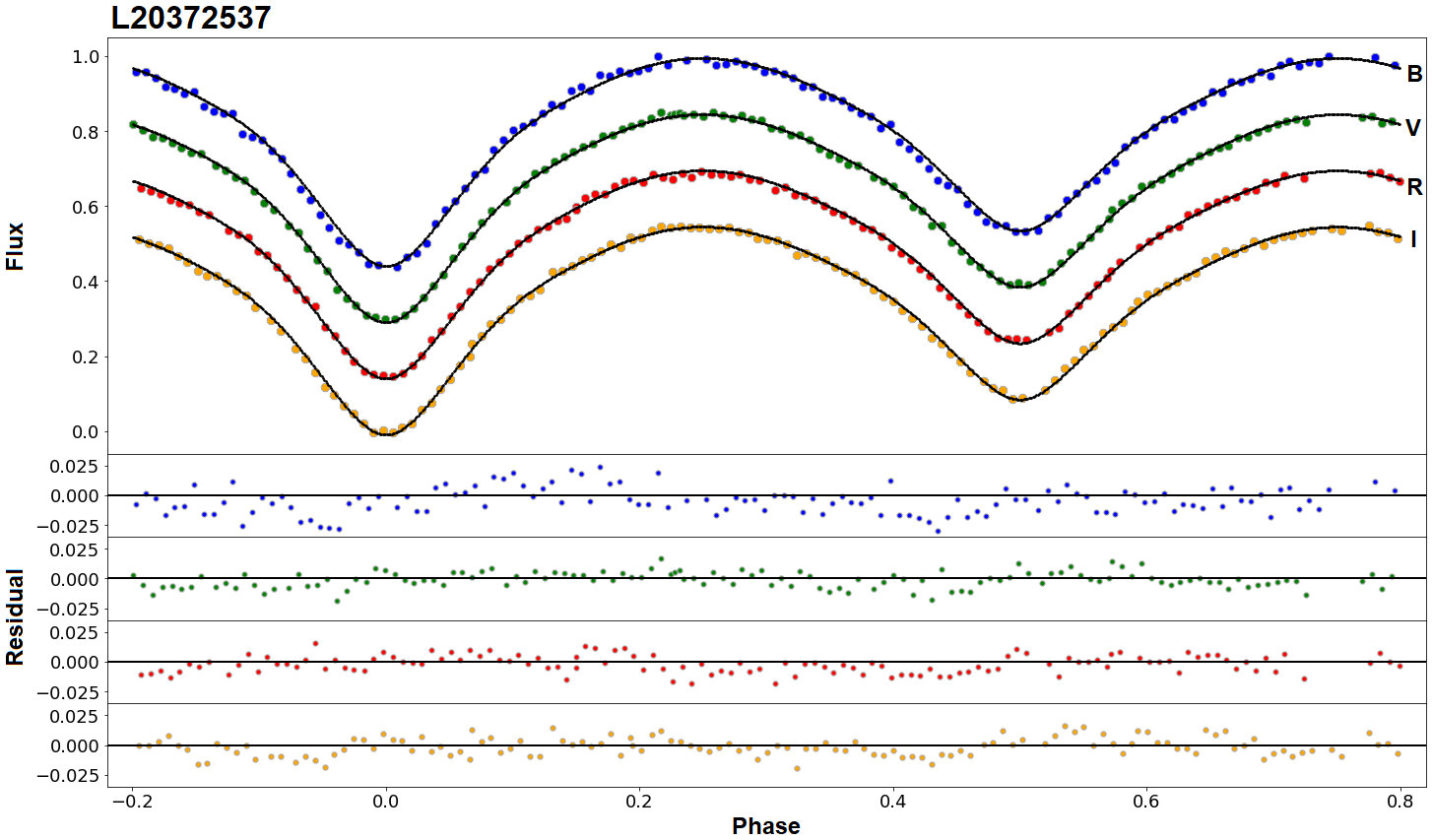}
\includegraphics[width=0.54\textwidth]{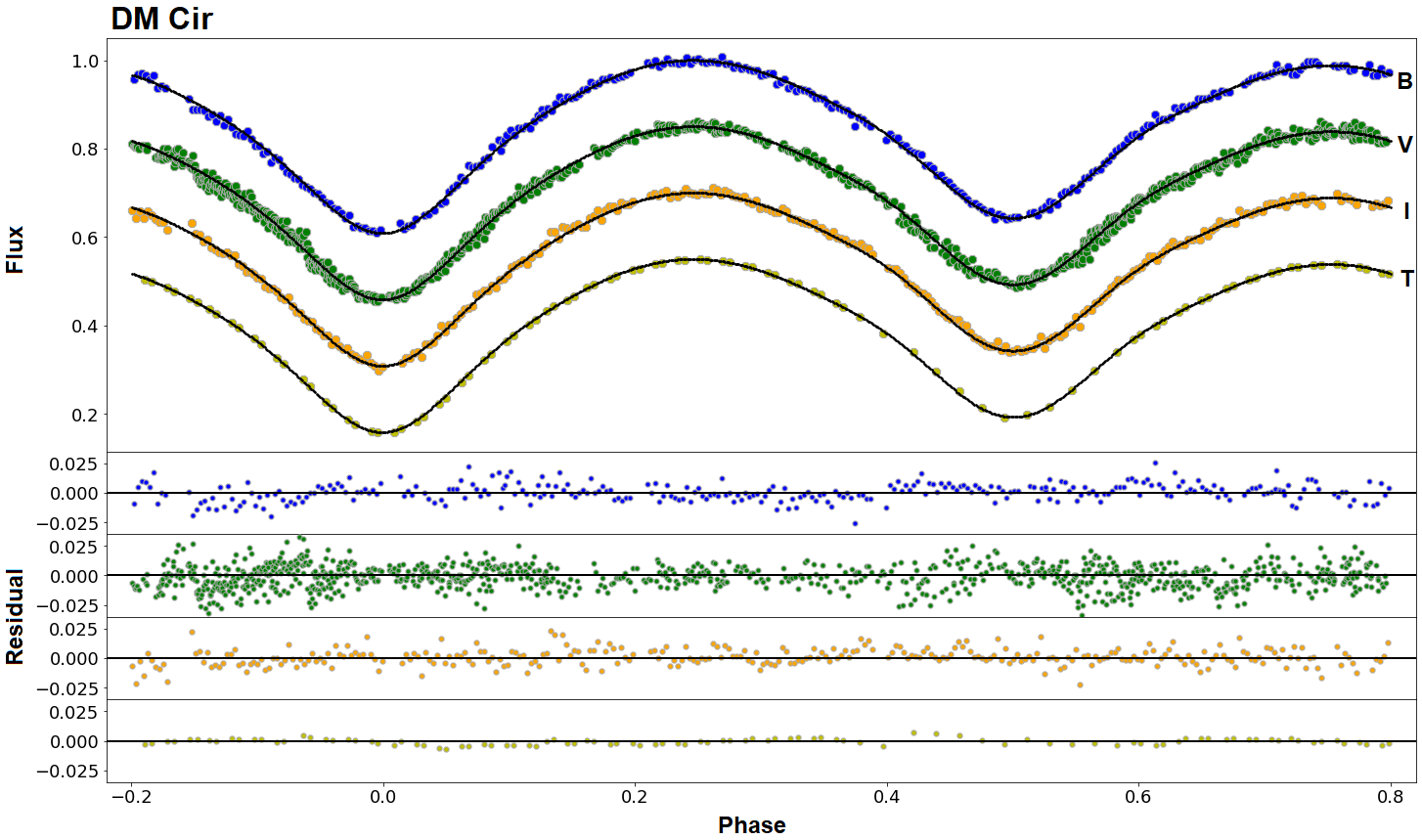}
\caption{The colored dots represent the observed light curves of the systems in different filters, and the synthetic light curves, generated using the light curve solutions, are also shown. Residuals are shown at the bottom of each panel.}
\label{fig-lc}
\end{figure*}

\begin{figure*}
\centering
\includegraphics[width=0.9\textwidth]{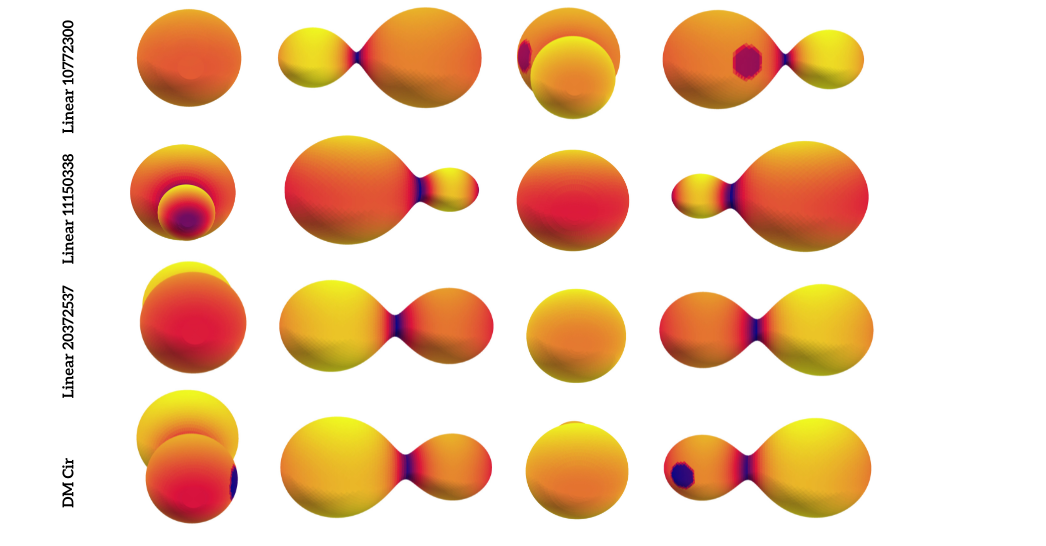}
\caption{3D view of the stellar components in the four target binary systems at orbital phases 0, 0.25, 0.50, and 0.75, respectively.}
\label{fig-3d}
\end{figure*}

The Figure \ref{fig-MCMC} shows the corner plots based on the heat-map for targets, providing a visual representation of parameter distributions from MCMC analysis. Table \ref{tab-solutions} presents the final results of the light curve modeling along with their uncertainties. Figure \ref{fig-lc} shows the observed and theoretical light curves in different filters. Additionally, Figure \ref{fig-3d} illustrates the three-dimensional (3D) structure of the binary systems and the starspots on the stars. The color in the figure represents the variations in effective temperature across the stellar surfaces.

\begin{table*}
\renewcommand\arraystretch{1.2}
\caption{Photometric solutions for the four target systems derived from light curve modeling.}
\centering
\begin{center}
\footnotesize
\begin{tabular}{c c c c c}
 \hline
Parameter & L10772300 & L11150338 & L20372537 & DM Cir \\
\hline
$T_{1}$ (K) & $5225(38)$ & $6155(33)$  & $4595(35)$ & $5770(32)$ \\
$T_{2}$ (K) & $5082(36)$ & $6185(29)$  & $4446(32)$ & $5622(31)$ \\
$q=M_2/M_1$ & $2.992(101)$ & $0.124(2)$ & $0.669(71)$ & $0.403(19)$ \\
$i^{\circ}$ & $81.63(67)$ & $80.63(36)$ & $82.86(77)$ & $73.21(51)$ \\
$f$ & $0.057(22)$ & $0.640(34)$ & $0.147(15)$ & $0.250(31)$ \\
$\Omega_1=\Omega_2$ & $ 6.570(311) $ & $1.979(34)$ & $3.131(259)$ & $2.622(35)$ \\
$l_1/l_{tot}$  & $0.295(20)$ & $0.857(18)$ & $0.643(36)$ & $0.719(6)$ \\
$l_2/l_{tot}$  & $0.705(20)$ & $0.143(18)$ & $0.357(36)$ & $0.281(6)$ \\
$r_{1(mean)}$ & $0.291(2)$ & $0.585(4)$ & $0.428(7)$ & $0.476(4)$ \\
$r_{2(mean)}$ & $0.480(1)$ & $0.246(4)$ & $0.357(7)$ & $0.319(4)$ \\
\hline
$Col._{spot}$(deg) & 95(1) &  &  & 92(1)  \\
$Long._{spot}$(deg) & 302(2) &  &  & 245(2)  \\
$Rad._{spot}$(deg) & 16(1) &  &  & 20(1)  \\
$T_{spot}/T_{star}$ & 0.92(1) &  &  & 0.90(1)  \\
Component & Secondary &  &  & Secondary \\
\hline
\end{tabular}
\end{center}
\label{tab-solutions}
\end{table*}

\vspace{0.6cm}
\section{Absolute Parameters}
\label{sec6}
One possible method for estimating the absolute parameters of contact binary systems, where only photometric data are available, is to use the Gaia DR3 parallax, as thoroughly discussed in the study by \cite{poro2024estimating}.
Estimating absolute parameters with the Gaia DR3 parallax method has certain limitations and requires an extinction coefficient ($A_V$) value lower than approximately 0.4 (\citealt{poro2024global}).
We estimated the $A_V$ for four systems based on the 3D dust-map Python package and Gaia DR3 distance data (\citealt{green20193d}). The results were in an acceptable state for using the Gaia DR3 parallax method to determine the absolute parameters of the stars.

The $V_{max}$ for each of the four systems was obtained from our ground-based observations and listed in Table \ref{tab-analysis}. Also, the bolometric correction ($BC$) was estimated using the method provided by the \cite{flower1996transformations} study.
Estimating the system's parameters requires several inputs, including the distance from Gaia DR3, $A_V$, $V_{max}$, $l_{1,2}/l_{tot}$, $BC_{1,2}$, $T_{1,2}$, $r_{mean_{1,2}}$, and $P$. With these parameters, we can determine $M_{V_{1,2}}$, $M_{bol_{1,2}}$, $R_{1,2}$, $L_{1,2}$, $a_{1,2}$, and $M_{1,2}$.
First, the method determined the absolute magnitude of the star ($M_V$) by using the apparent magnitude in the $\mathit{B V R_c I_c }$ filter, the distance from Gaia DR3 (d) in parsecs, and $A_V$. Subsequently, using the $l_{1,2}/l_{tot}$ parameter obtained from the standard filters in the light curve solutions process, $M_{V_{1,2}}$ are calculated. The bolometric correction was applied to derive the bolometric absolute magnitude ($M_{bol}$) for each star.

The radius of each star in the binary systems can be estimated through the relationship between $M_{bol}$ and luminosity ($L$). Once $L$ is derived and $T$ is known, the radius ($R$) of each star can then be estimated.
The semi-major axis $a(R_{\odot})$ of each system is then derived using $R_{1,2}$, $r_{mean1,2}$, and averaging $a_1(R_{\odot})$ and $a_2(R_{\odot})$.
The individual component masses can be determined by applying Kepler's third law, using the values of $a(R_{\odot})$, $P_{orb}$, and $q$, as formulated in Equations \ref{eqM1} and \ref{eqM2}.

\begin{equation}\label{eqM1}
M_1 = \frac{4 \pi^2 a^3}{G P^2 (1 + q)},
\end{equation}

\begin{equation}\label{eqM2}
M_2 = q \times M_1.
\end{equation}

Furthermore, the orbital angular momentum ($J_0$) of the systems was calculated using Equation \ref{eqJ0} from \cite{eker2006dynamical}, where $q$ represents the mass ratio, $M$ is the total system mass, $P$ is the orbital period, and $G$ is the gravitational constant.

\begin{equation}\label{eqJ0}
J_0=\frac{q}{(1+q)^2} \sqrt[3] {\frac{G^2}{2\pi}M^5P}.
\end{equation}

The estimated absolute parameters derived for the four target systems are listed in Table \ref{tab-analysis}.

\begin{table*}
\renewcommand\arraystretch{1.2}
\caption{Absolute parameter estimation for the four target systems using Gaia DR3 parallaxes and astrophysical relations.}
\centering
\begin{center}
\begin{tabular}{c c c c c}
\hline
Parameter & L10772300 & L11150338 & L20372537 & DM Cir \\
\hline
$M_1 (M_\odot)$ & 0.40(11) & 1.16(27) & 1.08(44) & 1.10(12)\\
$M_2 (M_\odot)$ & 1.20(31) & 0.14(3) & 0.73(30) & 0.44(5)\\
$R_1 (R_\odot)$ & 0.56(5) & 1.27(15) & 0.96(15) & 1.23(6)\\
$R_2 (R_\odot)$ & 0.93(9) & 0.51(3) &  0.80(11) & 0.82(4)\\
$L_1 (L_\odot)$ & 0.21(3) & 2.07(44) & 0.37(11) & 1.50(12)\\
$L_2 (L_\odot)$ & 0.52(9) & 0.34(4) & 0.23(5) & 0.61(4)\\
$M_{bol1}$(mag) & 6.45(14) & 3.95(23) & 5.83(31) & 4.30(8)\\
$M_{bol2}$(mag) & 5.46(19) & 5.90(11) & 6.36(26) & 5.29(7)\\
$\log g_1$(cgs) & 4.55(18) & 4.30(20) & 4.51(31) & 4.30(9)\\
$\log g_2$(cgs) & 4.58(20) & 4.18(15) & 4.49(29) & 4.26(8)\\
$a(R_\odot)$ & 1.92(17) & 2.12(16) & 2.24(30) & 2.58(9)\\
$logJ_0$(cgs) & 51.50(19) & 51.11(18) & 51.72(33) & 51.58(9)\\
\hline
$V_{max}(mag)$ & 15.93(10) & 15.51(11) & 16.48(7) & 10.11(8)\\
$A_V(mag)$ & 0.026(1) & 0.037(1) & 0.109(1) & 0.029(1)\\
$BC_1$ & -0.219(13) & -0.026(3) & -0.532(24) & -0.081(5)\\
$BC_2$ & -0.273(15) & -0.023(3) & -0.641(26) & -0.109(7)\\
\hline
\end{tabular}
\end{center}
\label{tab-analysis}
\end{table*}

\vspace{0.6cm}
\section{Discussion and Conclusion}
\label{sec7}
We present the first comprehensive analyses of light curves and estimations of the absolute parameters for four contact binary star systems. Ground-based observations were conducted at two observatories located in the Northern and Southern Hemispheres, utilizing multiband CCD photometric filters. Time-series photometric data from TESS were only available for the DM Cir system. We modeled and analyzed the orbital period variations and light curves using the PHOEBE Python code. An acceptable fit to the observations is reached under the cool-spot assumption, implying that a cool spot on the secondary component is needed to explain the asymmetry in the maxima. We estimated the absolute physical parameters of the four systems using the Gaia DR3 parallax method. From the overall investigations, we drew the following conclusions:

$\bullet$ We extracted the times of minima from our observations and supplemented them with minima from the different literature. We updated the ephemeris for the L10772300, L11150338, L20372537, and DM Cir systems. The observational time span for these systems is approximately 8, 9, 10, and 4 years, respectively. Due to the short observational time span between the earliest and latest observed minima, as well as the limited number of available minima in the O-C diagrams, a linear least-squares fit was considered as the best representation of the data.

$\bullet$ The light curve solutions indicate that the effective-temperature differences between the components of systems are small, with the lowest value of $30\,\mathrm{K}$ for L11150338 and the highest of $149\,\mathrm{K}$ for L20372537. This result suggests that the temperature differences between the components of the target systems are consistent with the range of contact binaries (\citealt{poro2025IV}).
The effective-temperature difference, defined as $\Delta T=|T_{1}-T_{2}|$, quantifies the thermal disequilibrium between the two stellar components. In contact binary systems, both stars fill their Roche lobes and are enclosed within a common convective envelope, sharing a single equipotential surface. This configuration allows for continuous energy exchange through the common envelope, leading to thermal equilibrium between the components.
From another point of view, according to \cite{webbink2003contact}, the nearly equal depths of primary and secondary eclipses, as well as the absence of pronounced color variations in W UMa-type light curves, suggest that the components must be nearly identical in effective temperature. Detailed light curve synthesis models employing Roche geometry further confirm that the components of such binaries exist in a state of physical contact (\citealt{webbink2003contact}).
The near-uniformity of effective temperature over the surface of a W UMa binary arises naturally from hydrostatic equilibrium within the common envelope, and that energy exchange between contact components is thus implicitly demanded by hydrostatic equilibrium. This interpretation is consistent with the theoretical framework proposed by \cite{lucy1968structure}, in which the continuity of entropy within the common envelope enforces nearly equal fluxes from both stars. Consequently, the observed near equality of the component temperatures in our systems reflects energy transfer and thermal contact.
Moreover, such near-equal effective temperatures are consistent with the behavior of most W UMa-type binaries reported in the literature, such as \cite{poro2025bsn}, where the majority of targets exhibit only small temperature differences between their components. The derived temperature difference of less than $400\,\mathrm{K}$ supports the interpretation that the components of our targets are in, or very close to, surface thermal equilibrium, while internal thermal equilibrium is not necessarily expected (\citealt{li2004structure,yakut2005evolution,2025ApJ...995...19F}).
The spectral classifications of the component stars were determined using the \cite{cox2015allen} and \cite{eker2006dynamical} studies (Table \ref{tab-conclusion}).

$\bullet$ We employed the Gaia DR3 parallax method to estimate the absolute parameters of each system. This method is well-suited for this purpose, providing geometrically determined distances with substantially reduced systematic errors. The zero-point calibration by \cite{lindegren2021gaia}, based on quasars, Large Magellanic Cloud stars, and physical binaries, guarantees consistent and precise measurements. However, this approach is reliable when the $A_V$ value is low (\citealt{poro2024global}). Our target systems align with this circumstance, as shown in Table \ref{tab-analysis}. 
In the Gaia DR3 parallax approach, the total orbital semi-major axis is independently inferred from the properties of each component, yielding two determinations, $a_1$ and $a_2$. In a physically consistent solution, these values are expected to agree; therefore, their difference, $\Delta a = |a_1 - a_2|$, is used as an internal consistency check between the astrometric parallax, orbital geometry, and the derived mass ratio (\citealt{hilditch2001introduction,2025d}). A $\Delta a < 0.1$ indicates acceptable consistency and serves as an additional proof of the reliability of the light curve solutions (\citealt{poro2024estimating}). As presented in Table \ref{tab-conclusion}, the $\Delta a$ values obtained for the targets are less than 0.1. Additionally, based on the effective temperature and mass of each component, it was determined that L10772300 and L11150338 are W-subtype, while the others are A-subtype.

\begin{figure*}
\centering
\includegraphics[width=0.44\textwidth]{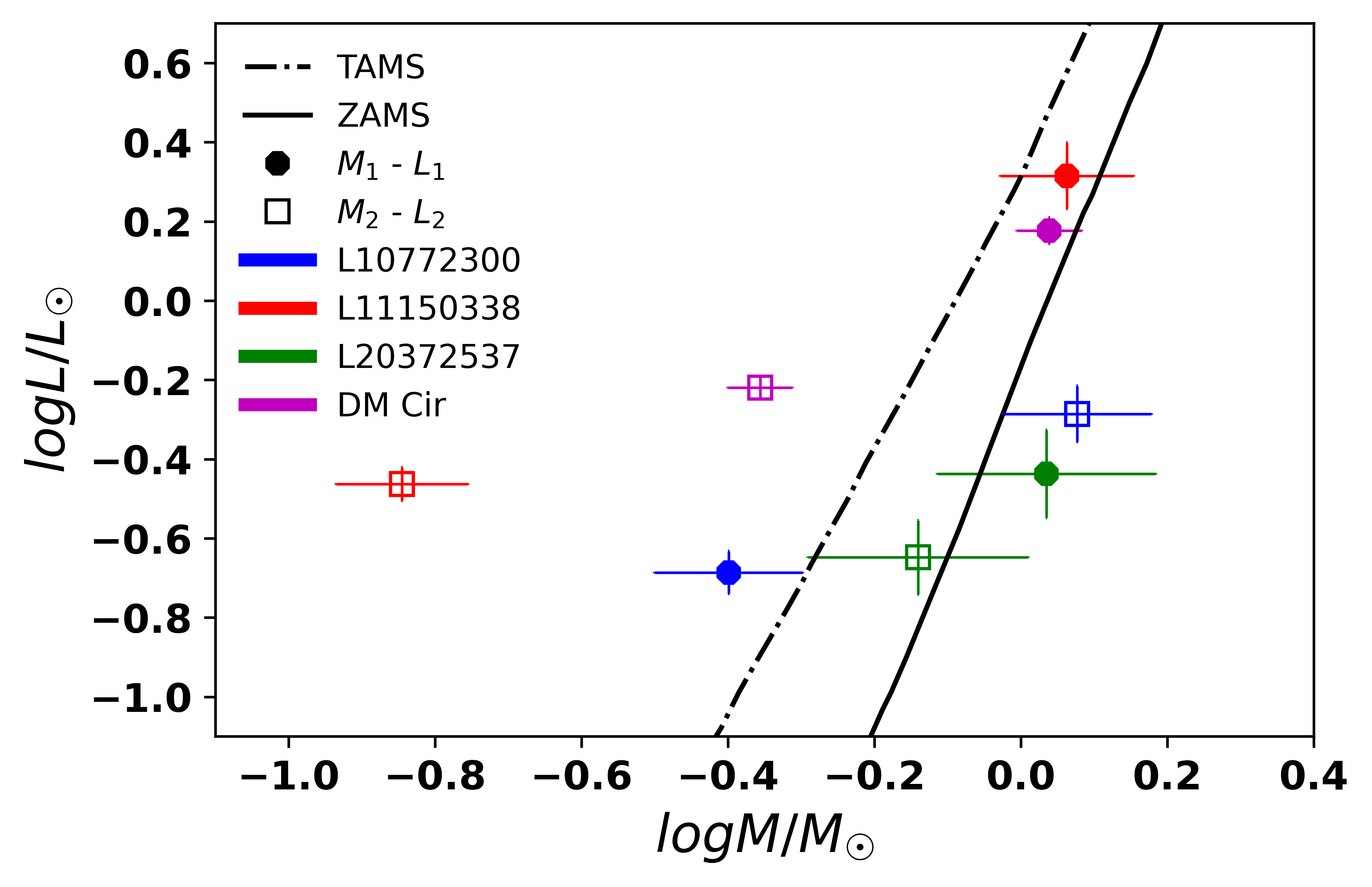}
\includegraphics[width=0.44\textwidth]{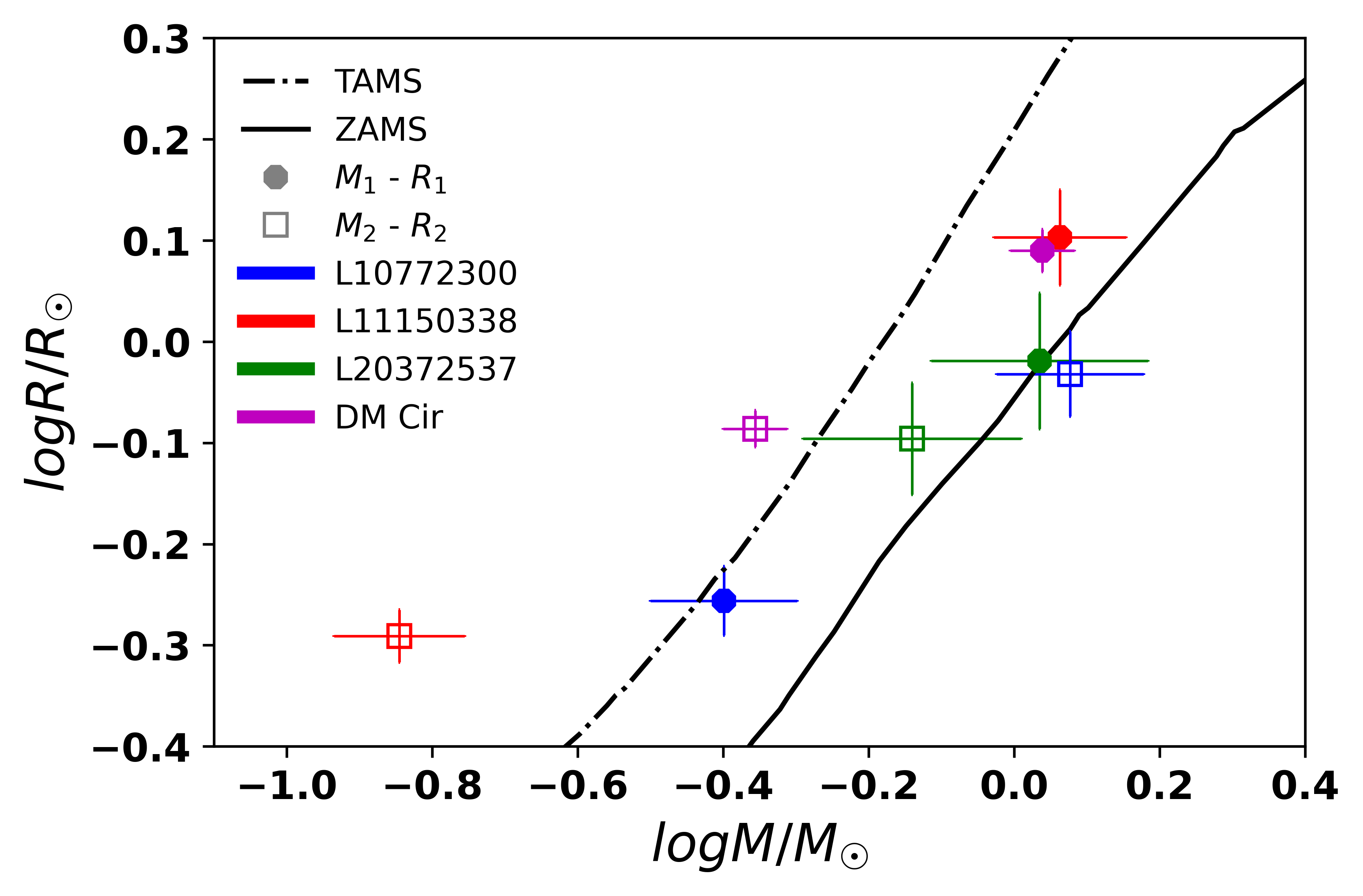}
\includegraphics[width=0.44\textwidth]{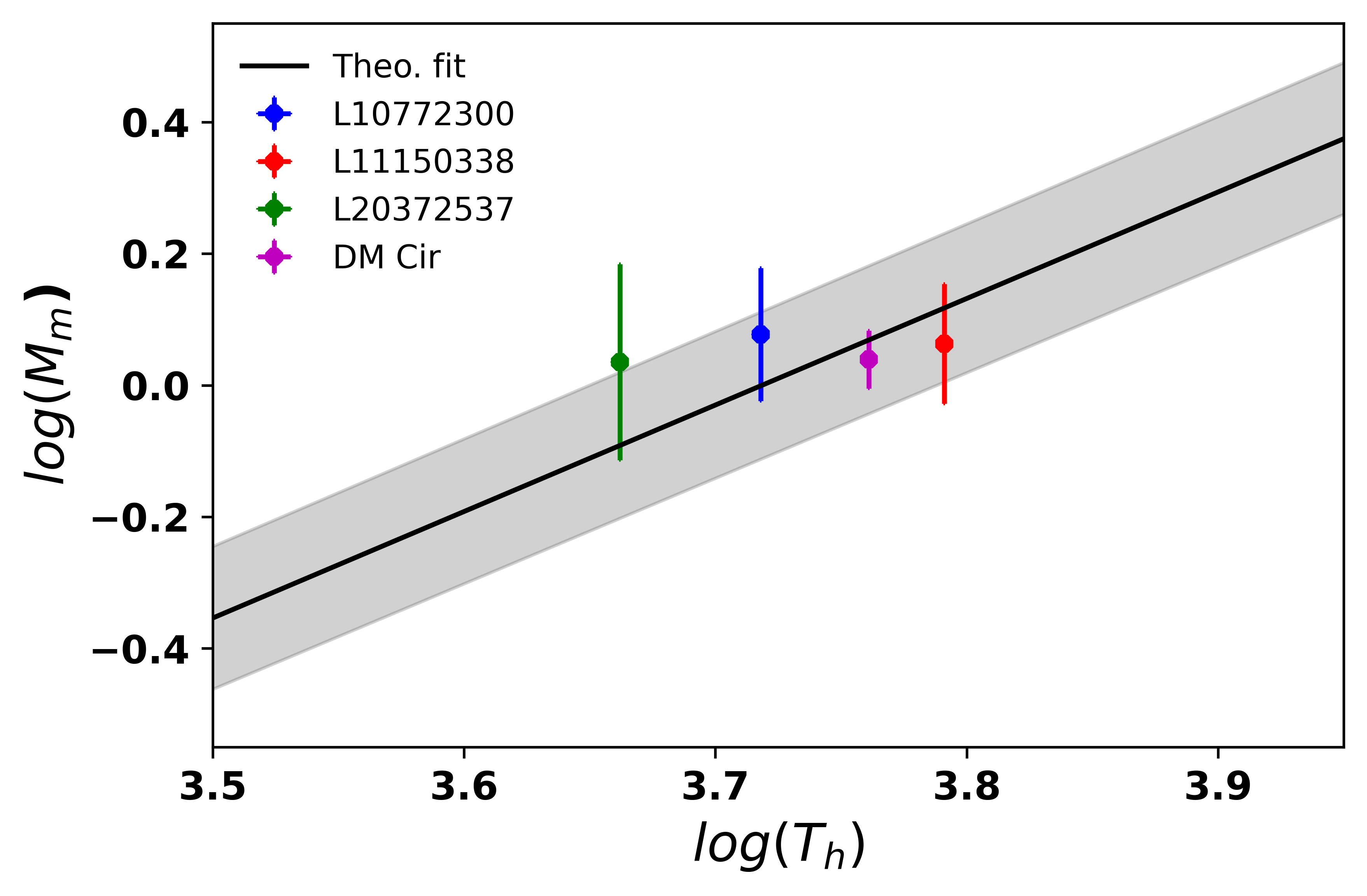}
\includegraphics[width=0.44\textwidth]{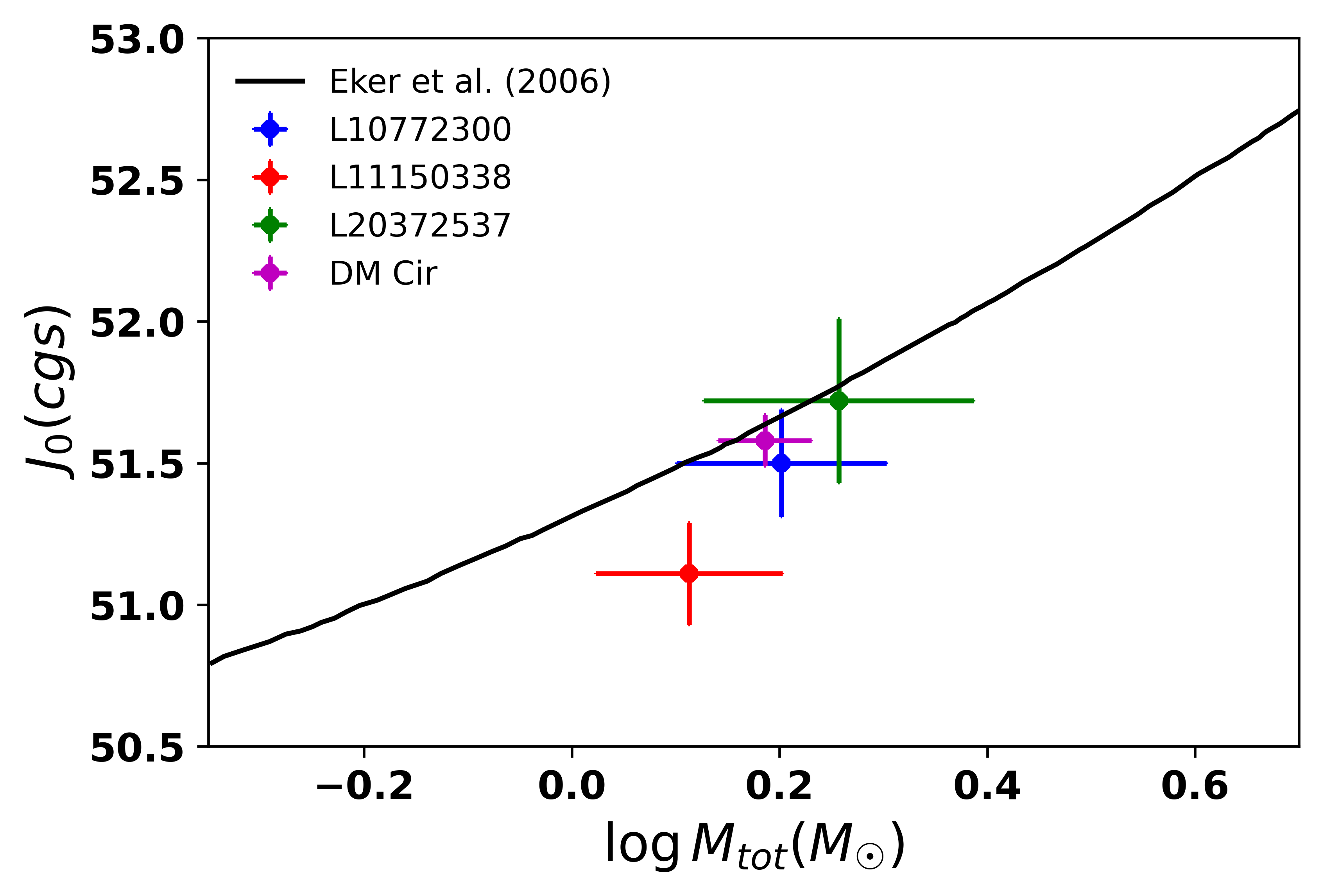}
\caption{$M$–$L$, $M$–$R$, $T_h$–$M_m$, and $M_\mathrm{tot}$–$J_0$ relation diagrams for the studied contact binaries.}
\label{fig-MLRJ0T}
\end{figure*}

$\bullet$ To investigate the evolutionary status of the four target systems, we constructed the mass–luminosity ($M-L$) and mass–radius ($M-R$) diagrams using the derived absolute parameters (Figure \ref{fig-MLRJ0T}). The theoretical Zero-Age Main Sequence (ZAMS) and Terminal-Age Main Sequence (TAMS) relations were taken from \cite{hurley2002evolution} and are plotted as solid and dashed black lines, respectively. As shown in Figure \ref{fig-MLRJ0T}, the $M-L$ and $M-R$ diagrams indicate that more mass stars are generally positioned closer to the ZAMS, while lower-mass components tend to lie near the TAMS. However, there is one exceptional target in this study; the lower-mass component of L11150338 lies away from the TAMS.
This behavior is consistent with the long-standing interpretation that contact binaries, as interacting systems, cannot in general satisfy single-star mass-radius relations. As pointed out by \cite{kuiper1941interpretation}, such relations are only expected to hold in the limiting case of equal-mass components, while systems with unequal masses necessarily deviate from single-star expectations. Subsequent studies have shown that binary interaction phases modify the internal structure of the components, altering the envelope-to-core mass ratio and, consequently, their mass-radius relations (\citealt{yakut2005evolution, stepien2005evolutionary}). In such systems, the less massive component is typically a more evolved star with a lower central hydrogen content and an expanded radius, while the more massive component lies close to the main sequence. If the less massive component were to evolve significantly beyond the main sequence, the system would likely merge on a thermal timescale (\citealt{poro2026detailed}). This structure results from angular-momentum loss–driven evolution and past mass exchange leading to a mass-ratio reversal. 
Therefore, any direct comparison with the ZAMS and TAMS lines for single stars should be made with caution.
Following the analysis of \cite{poro2024first}, the temperature-mass ($T_h-M_m$) relationship for contact binary systems is presented in Figure \ref{fig-MLRJ0T}, where a linear relation (Equation \ref{Mm}) is fitted to the data. 
Here $M_m$ denotes the mass of the more massive component.

\begin{equation}\label{Mm}
\log M_{m} = (1.6185 \pm 0.0150) \times (\log T_{h}) + (-6.0186 \pm 0.0562).
\end{equation}

The positions of the higher-mass, hotter components of each target system on the $T_h-M_m$ diagram align with the distribution of contact binaries. These components lie within the typical region of contact binary systems. We estimated the orbital angular momentum for each system (Table \ref{tab-analysis}), and their locations are shown on the \( J_{0} - \log M_{tot} \) diagram (Figure \ref{fig-MLRJ0T}). We utilized the parabolic line from the study by \cite{eker2006dynamical}, which shows that our targets are located below this curve, within the domain typically associated with contact binaries.

Based on the results of this study, we computed the ratio of spin to orbital angular momentum ($J_{\rm spin}/J_{\rm orb}$) for the targets using the classical relation

\begin{equation}
\frac{J_{\rm spin}}{J_{\rm orb}} = \frac{1+q}{q} \Big[ (k_1 r_1)^2 + (k_2 r_2)^2 q \Big],
\end{equation}
where $q$ is the mass ratio, $r_1$ and $r_2$ are the fractional radii of the primary and secondary, and $k_1$ and $k_2$ are the respective gyration radii, with adopted values $k_1 = k_2 = 0.06$ from \cite{2006MNRAS.369.2001L}. We find that L10772300 has $J_{\rm spin}/J_{\rm orb} = 0.062(4)$, L11150338 has $0.190(4)$, L20372537 has $0.040(6)$, and DM Cir has $0.056(4)$. All of these values are well below the Darwin stability threshold ($J_{\rm spin}/J_{\rm orb} \sim 0.333$; \citealt{hut1980stability}), confirming that the systems are dynamically stable.

$\bullet$ In order to investigate the formation and evolution of W UMa-type contact binaries, \cite{yildiz2013origin} proposes a method based on the observational properties of overluminous secondary components. In their method, if the central physical condition of the system's secondary component is similar to a single star of mass $M_L\approx L^{0.25}$ (in solar units), the initial mass of the secondary component can be estimated from the following equation:
\begin{equation}\label{M2i}
M_{2i} = M_{2} + \Delta M = M_{2} + 2.50 \left( M_{L} - M_{2} - 0.07 \right)^{0.64},
\end{equation}
where $M_{2}$ represents the current mass of the secondary, and $M_{L}$ is the luminosity-based mass estimated using equation \ref{ML}.
\begin{equation}\label{ML}
M_{L} = \left( \frac{L_{2}}{1.49} \right)^{\frac{1}{4.216}}.
\end{equation}

The initial mass of the primary component was estimated using equation \ref{M1i}. In this approach, the mass lost from the system is expressed as $M_{\rm lost} = \gamma \,\Delta M$, where we adopted $\gamma = 0.664$ following the prescription of \cite{yildiz2013origin}. We emphasize that the degree of conservativeness of mass transfer depends on the evolutionary state of the mass-donor star. In particular, the value of $\gamma$ derived by \cite{yildiz2013origin} is based on the assumption that the initial mass transfer occurs when the donor is close to the TAMS. This assumption may not be strictly valid for long-lived contact binaries, which are expected to experience mass transfer during the main-sequence phase. Therefore, the initial mass estimates presented here should be regarded as model-dependent and subject to these underlying assumptions

\begin{equation}\label{M1i}
M_{1i} = M_{1} - (\Delta M - M_{\text{lost}}) = M_{1} - \Delta M (1 - \gamma).
\end{equation}

The formation and evolution of W UMa-type contact binaries are driven by nuclear and angular momentum evolution via magnetic braking (\citealt{mestel1968magnetic}), which causes the stars to fill their Roche lobes and exchange mass (\citealt{okamoto1970formation, vilhu1982detached}).
The initially more massive component transfers material to its companion, leading to a mass ratio reversal. This process explains the systems' nearly identical temperatures and the overluminosity of the secondary (\citealt{webbink2003contact, li2008evolutionary}), a remnant of its higher initial mass and the energy transfer between components within the shared envelope.

According to the evolutionary framework of \cite{tutukov2004evolutionary}, systems with $M_{1\rm i}$ between $0.2$ and $1.5\,M_\odot$ are expected to experience angular momentum loss. This process leads to orbital contraction and can ultimately result in contact between the components. As shown in Table \ref{tab-conclusion}, our systems fall within this mass range. However, the efficiency of magnetic braking is known to decrease for higher-mass primaries and becomes significantly reduced for $M_{1\rm i} \gtrsim 1.3\,M_\odot$ (e.g. \citealt{1988A&A...191...57P,2002ApJ...565.1107P}). Therefore, the evolution of our systems is most likely governed by magnetic braking, which drives the loss of orbital angular momentum and promotes their evolution toward contact configurations. The values of $M_{2\rm i}$ in our target systems are higher than their current masses (see Table \ref{tab-conclusion}), suggesting that the secondary components of these contact binaries have internal structures that might differ from those of normal main-sequence stars (\citealt{yildiz2013origin}).

The difference between the current secondary masses and the inferred initial values $M_{2\rm i}$ reflects the assumptions adopted in our evolutionary framework, particularly the non-conservative mass transfer prescription with $\gamma = 0.664$ following \cite{yildiz2013origin}. Given the masses and orbital periods, target systems are most plausibly formed through AML, with the onset of interaction occurring during the main-sequence phase. In such systems, both AML and Roche-lobe overflow (RLOF) are expected to operate concurrently during their evolution toward contact configurations (\citealt{hilditch1988evolutionary,eggleton2012formation}). Compared to the other targets, the mass lost in L20372537 is lower and can be explained by its higher mass ratio relative to the other three systems.

$\bullet$ Energy is transferred from the hotter component to the cooler one through the common envelope and causes the two components to have nearly the same temperature even though their masses are quite different (\citealt{lucy1968structure,lucy1968light}). Building on this behavior, understanding the mechanisms and efficiency of energy transfer has been a longstanding focus of extensive studies. A significant step forward was provided by \cite{mochnacki1981contact}, who showed the role of mass ratio in the relative energy transfer rate and normal mass–luminosity relation for independent stars. Subsequent studies, such as \cite{wang1994thermal}, further expand this framework and demonstrate that the relative energy transfer rate increases continuously as the mass ratio increases, while \cite{liu2000relation} showed that it also depends on the evolutionary degree of the primary. In addition, \cite{csizmadia2004properties} analyzed the properties of these systems using the large catalog of contact binaries to show that the energy transfer parameter can be expressed as a simple function of the mass and luminosity ratios.

Building on these foundations, \cite{jiang2009energy} analyzed a sample of 133 W UMa–type contact binaries and examined how energy transfer impacts the secondary components. Based on the distribution of temperature differences, they showed that properties of the common envelope might not have a significant effect on energy transfer between the components of these systems. Under the assumptions of contact configurations with nearly uniform effective temperatures and ZAMS primaries, the relative energy transfer rates can be described using the formalism introduced by \cite{mochnacki1981contact}

\begin{equation}\label{Eq.U1}
U_1 = \frac{r^{2} t^{4} - q^{\alpha}}{1 + q^{\alpha}}
\end{equation}

\begin{equation}\label{Eq.U2}
U_2 = \log [\frac{r^{2} t^{4} - q^{\alpha}}{q^{\alpha} (1 + r^{2} t^{4})}],
\end{equation}
where $r={R_{2}}/{R_{1}}$, $t={T_{2}}/{T_{1}}$ correspond to radius and temperature ratios, and $\alpha = 3.42$ represents the exponent of the mass luminosity relation.

In order to investigate the overluminosity of secondaries and its connection to energy transfer between the components, the nuclear luminosity of each component can be approximated using the relation given by \cite{demircan1991stellar}

\begin{equation}\label{Eq.L0}
L_{0} \simeq 1.03\, M^{3.42}, \qquad 0.1 \le M \le 120,
\end{equation}

Equation (\ref{Eq.L0}) represents an approximate mass-luminosity relation for single main-sequence stars and should therefore be applied and interpreted with caution for contact binary components.

\begin{figure*}
\centering
\includegraphics[width=0.99\textwidth]{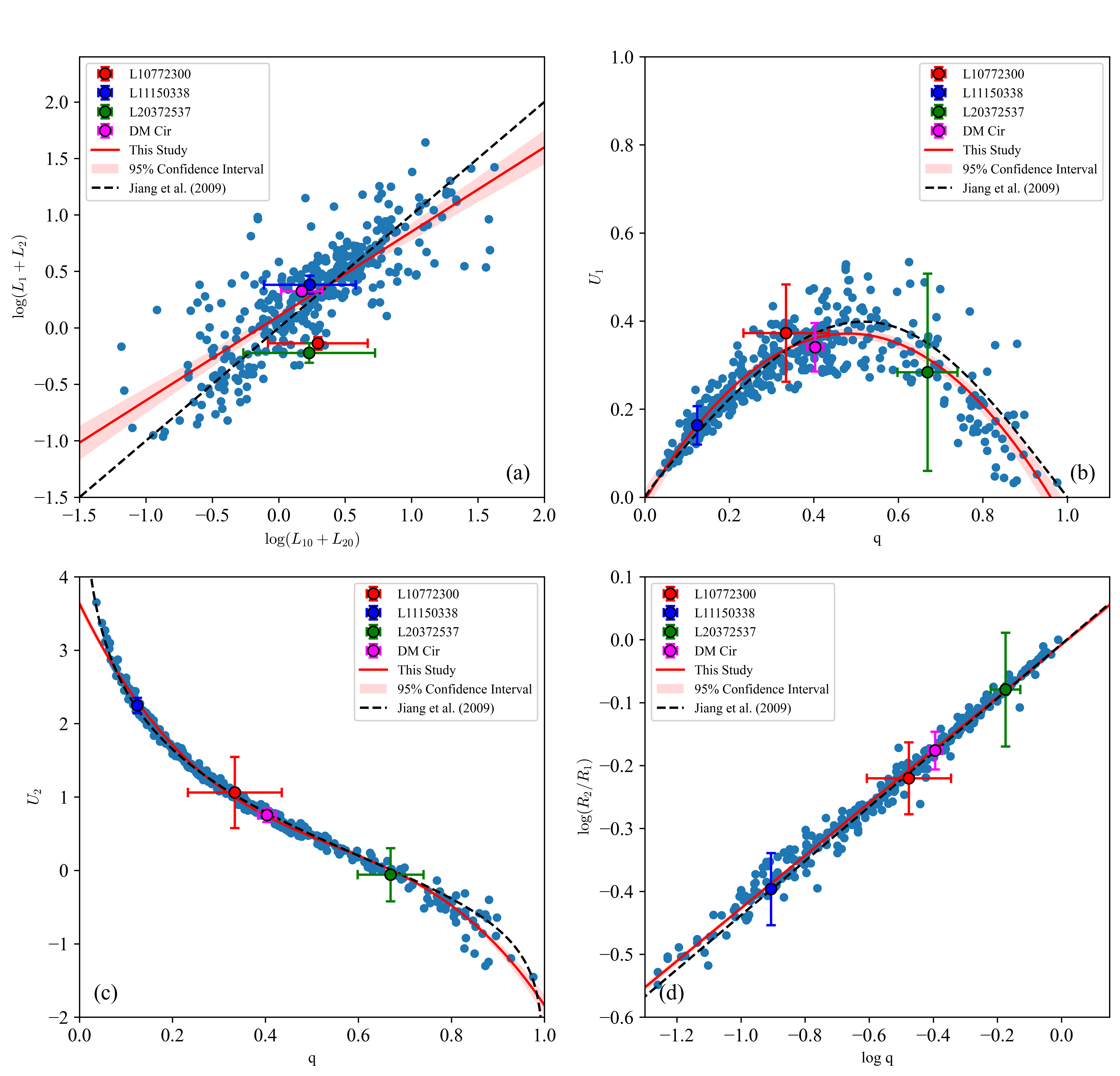}
\caption{Distributions of $\log(L_{1}+L_{2})$ versus $\log(L_{10}+L_{20})$, $U_{1}$, $U_{2}$, and $\log(R_{2}/R_{1})$ as functions of mass ratio for 411 W UMa systems (blue dots)and the solid lines represent the statistical fits.}
\label{fig_TE}
\end{figure*}

As shown in Figure \ref{fig_TE}, \cite{jiang2009energy} derived the theoretical luminosity relation in which $\log(L_{1} + L_{2})=\log(L_{10} + L_{20})$, along with theoretical fits for the distribution of $U_1$ and $U_2$ as a function of q, as well as $\log(R_{2}/R_{1})$ versus $\log q$, expressed as

\begin{equation}\label{Eq.theorycurve1}
U_{1} = \frac{q^{0.92}-q^{3.42}}{1 + q^{3.42}},
\end{equation}

\begin{equation}\label{Eq.theorycurve2}
U_{2} = \log [\frac{q^{0.92} - q^{3.42}}{q^{3.42}(1 + q^{0.92})}].
\end{equation}

\begin{equation}\label{Eq.theoryfit4}
\log({R_2}/{R_1})
= (0.431\pm 0.006)\log q-(0.007\pm 0.003).
\end{equation}

Equation \ref{Eq.theorycurve1} closely reproduces the results of \cite{2023A&A...672A.175F} for the fillout factor when the Roche-lobe areas of the secondary and primary stars, $S_2$ and $S_1$, are approximated to be proportional to their masses, that is, $S_2/S_1 \simeq q$. Relation \ref{Eq.theoryfit4} is consistent with the classical Roche-lobe geometry and the well-established relations originally described by \cite{kopal1959close}.

To extend the applicability of this theoretical framework, we have collected the physical parameters of 407 W UMa-type binaries from the study by \cite{poro2025study}, along with the four target systems analyzed in this study. Although the catalog in that study contains 818 systems, we restricted our sample to those that are consistent with the assumptions provided by \cite{jiang2009energy}. We considered only systems in which both components fill their inner Roche lobes (${R_2}/{R_1} = q^{0.46}$), and have a $\Delta T_{{eff}} < 400{K}$. In addition, the primaries were required to be consistent with zero-age main sequence stars. It is also important to note that systems lacking any of the required parameters are excluded from the sample, as these values are essential for our analysis.
Using this selective dataset, we calculated the energy transfer rate of the primary ($U_{1}$) and secondary ($U_{2}$) components for a wide range of mass ratios and estimated the nuclear luminosity for each component. 
Figure \ref{fig_TE} shows the full sample of 411 targets, Table \ref{tab-TE} lists the results for the four systems analyzed in detail, and statistical fits (with uncertainties) were performed for each panel using equations \ref{Eq.fit1}–\ref{Eq.fit4}.

First, the relation between $\log(L_{1} + L_{2})$ and $\log(L_{10} + L_{20})$ for both A- and W-subtype binaries is well described by a statistical linear fit.

\begin{equation}\label{Eq.fit1}
\log(L_{1} + L_{2})=(0.747\pm 0.083)\,\log(L_{10} + L_{20}) + (0.104\pm 0.047).
\end{equation}

The best fit for the relative energy transfer rates versus the mass ratio is represented by a second-degree polynomial and a third-degree polynomial, given by

\begin{equation}\label{Eq.fit2}
U_{1}=(-1.624\pm 0.149)\,q^{2}+(1.567\pm 0.139)\,q - (0.006\pm0.027).
\end{equation}

\begin{equation}\label{Eq.fit3}
U_{2} = (-11.368\pm 1.276)\,q^{3}+(18.814\pm 1.780)\,q^{2}-(12.914\pm 0.713)\,q+(3.632\pm 0.080).
\end{equation}

The relation between $\log({R_{2}}/{R_{1}})$ and $\log(q)$ is well described by a logarithmic fit.

\begin{equation}\label{Eq.fit4}
\log\!\left(\frac{R_{2}}{R_{1}}\right)=(0.419\pm 0.007)\,\log(q) - (0.008\pm 0.004).
\end{equation}

As shown in Figure \ref{fig_TE}, the 411 systems align well with the theoretical relations derived by \citet{jiang2009energy} in all four panels.

L10772300 located below both the statistical fit and the theoretical luminosity relation in panel (a), exhibiting a deviation of $|\Delta \log L|>0.4$. Such deviations often indicate evolved components, particularly in low-mass W UMa binaries, which exhibit higher velocity dispersions and older mean kinematic ages (\citealt{bilir2005kinematics}). Alternatively, they may arise from the presence of a third component whose light contribution biases the light curve solution and shifts the inferred luminosities (\citealt{pribulla2006contact}).

In panel (b), L10772300 is near the peak of the $U_{1}(q)$ curve, indicating that the primary must transfer the maximum energy to bring the secondary to the same temperature. In panel (c), its $U_{2}$ value confirms that the secondary is overluminous, the expected consequence of strong thermal coupling in contact binaries (\citealt{webbink2003contact}). We emphasize that, in our sample, the secondary is defined as the less massive component, regardless of its temperature, following the convention that the primary is always the more massive star in contact binaries.

L11150338 expresses strong energy transfer, distinguishing it from the other target systems. In panel (a), it is above the statistical and theoretical lines, indicating that its observed luminosity exceeds its nuclear luminosity, suggesting thermal redistribution. L11150338 has a low mass ratio, corresponding to a low $U_{1}$. Panel (c) shows that L11150338 exhibits the largest $U_{2}$ value among the four targets, indicating that its secondary is strongly overluminous relative to its nuclear luminosity, which is expected due to its relatively high mass ratio.

L20372537 is located below both the statistical and the theoretical luminosity relation in panel (a), again showing a deviation of $|\Delta \log L| > 0.4$. In panel (b), the mass ratio of L20372537 ($q=0.669$) places it in the descending part of $U_{1}(q)$, where the primary needs to transfer relatively less energy due to the comparable mass of the secondary. Its deviation from the curves is consistent with \cite{jiang2009energy}, which found that systems with larger unequal component temperatures display more deviation from the expected energy transfer curve. The value of $U_{2}$ for L20372537 is $-0.058$, indicating minimal luminosity enhancement of the secondary component.

DM Cir agrees well with both theoretical and statistical relations in all four panels, following the expectations for efficient energy transfer given the component radii, luminosities, and masses. Its position in parameter space suggests structural and thermal properties consistent with those expected for moderate mass ratio systems in stable contact.

Finally, all four target systems align closely with the theoretical relation from equation (8) of \citet{jiang2009energy} in panel (d), indicating that their geometry is consistent with a Roche filling configuration.

\begin{table*}
\caption{Some conclusions regarding to the evolutionary history of the four target systems.}
\centering
\begin{center}
\footnotesize
\begin{tabular}{c c c c c c c c}
\hline
Systems	& $|\Delta T|$ (K) & $|\Delta a|$ & Sp. cat. & Subtype & $M_{1i}(M_\odot)$ & $M_{2i}(M_\odot)$ & $M_{lost}(M_\odot)$ \\
\hline
L10772300 & 143 & 0.03 & K0-K1 & W & 0.942(331) & 1.168(241) & 0.510(231)\\
L11150338 & 30 & 0.08 & F8-F8 & W & 0.625(271) & 1.733(52) & 1.058(49)\\
L20372537 & 149 & 0.01 & K3-K2 & A & 1.016(188) & 0.921(69) & 0.127(11)\\
DM Cir & 148 & 0.01 & G5-G6 & A & 0.712(127) & 1.595(80) & 0.767(85)\\
\hline
\end{tabular}
\end{center}
\label{tab-conclusion}
\end{table*}

\begin{table*}
\caption{Relative energy transfer rates and nuclear luminosities for the components of the four target systems.}
\centering
\begin{center}
\footnotesize
\begin{tabular}{c c c c c}
\hline
Systems	& $L_{10}$ & $L_{20}$ & $U_{1,{rtq}}$ & $U_{2,{rtq}}$\\
\hline
L10772300 & 1.921(1698) & 0.045(42) & 0.373(110) & 1.063(485) \\
L11150338 & 1.711(1362) & 0.001(1) & 0.163(44) & 2.249(102)\\
L20372537 & 1.340(1867) & 0.351(493) & 0.284(224) & -0.058(362)\\
DM Cir & 1.427(532) & 0.062(24) & 0.340(55) & 0.755(94)\\
\hline
\end{tabular}
\end{center}
\label{tab-TE}
\end{table*}

\vspace{0.6cm}
\section*{Data availability}
The ground-based observations are provided in the online supplementary materials of this paper.

\vspace{0.6cm}
\section*{Acknowledgments}
We sincerely thank the referee for their constructive comments and valuable suggestions, which significantly improved the quality and clarity of this manuscript. This manuscript, including the observation, analysis, and writing processes, was provided by the BSN project ({\url{https://bsnp.info}}). Ground-based observations (LINEAR 10772300, LINEAR 11150338, and LINEAR 20372537 systems) were conducted with the cooperation of the Observatorio Astron\'omico Nacional on the Sierra San Pedro M\'artir (OAN-SPM), Baja California, M\'exico. We used IRAF, distributed by the National Optical Observatories and operated by the Association of Universities for Research in Astronomy, Inc., under a cooperative agreement with the National Science Foundation. We used results reported by the European Space Agency mission Gaia(\url{http://www.cosmos.esa.int/gaia}). In this work, we utilize observations obtained by the TESS mission, supported through NASA's Explorer Program.

\vspace{0.6cm}
\section*{Appendix}
The appendix provides the measured primary and secondary eclipse timings for the four analyzed contact binary systems, obtained from TESS photometric observations.

\renewcommand{\thetable}{A\arabic{table}}
\setcounter{table}{0}
\begin{table*}
\renewcommand\arraystretch{0.77}
\caption{Times of Minima extracted from the TESS data for DM Cir. The minimums have been reduced from 2,400,000.}
\centering
\begin{center}
\begin{tabular}{c c c c c c c c c c c c}
\hline
Min.(BJD) & Error & Epoch & O-C & Min.(BJD) & Error & Epoch & O-C & Min. & Error & Epoch & O-C\\
\hline
59333.8952	&	0.0002	&	16345.5	&	-0.0200	&	59346.2723	&	0.0001	&	16377.5	&	-0.0197	&	59359.8097	&	0.0002	&	16412.5	&	-0.0196	\\
59334.0885	&	0.0002	&	16346	&	-0.0201	&	59347.4328	&	0.0002	&	16380.5	&	-0.0196	&	59360.0023	&	0.0002	&	16413	&	-0.0203	\\
59334.2818	&	0.0002	&	16346.5	&	-0.0202	&	59347.6253	&	0.0002	&	16381	&	-0.0205	&	59360.1965	&	0.0002	&	16413.5	&	-0.0195	\\
59334.4751	&	0.0002	&	16347	&	-0.0203	&	59347.8196	&	0.0002	&	16381.5	&	-0.0196	&	59360.3892	&	0.0002	&	16414	&	-0.0203	\\
59334.6686	&	0.0002	&	16347.5	&	-0.0201	&	59348.0122	&	0.0002	&	16382	&	-0.0204	&	59361.9359	&	0.0002	&	16418	&	-0.0206	\\
59334.8618	&	0.0001	&	16348	&	-0.0204	&	59348.2064	&	0.0002	&	16382.5	&	-0.0196	&	59362.1305	&	0.0002	&	16418.5	&	-0.0194	\\
59335.0554	&	0.0002	&	16348.5	&	-0.0201	&	59348.3989	&	0.0002	&	16383	&	-0.0204	&	59362.3227	&	0.0002	&	16419	&	-0.0206	\\
59335.2486	&	0.0002	&	16349	&	-0.0204	&	59348.5931	&	0.0002	&	16383.5	&	-0.0196	&	59362.5173	&	0.0002	&	16419.5	&	-0.0194	\\
59335.4423	&	0.0002	&	16349.5	&	-0.0200	&	59348.7857	&	0.0001	&	16384	&	-0.0205	&	59362.7095	&	0.0002	&	16420	&	-0.0206	\\
59335.6353	&	0.0002	&	16350	&	-0.0204	&	59348.9799	&	0.0002	&	16384.5	&	-0.0196	&	59362.9041	&	0.0002	&	16420.5	&	-0.0194	\\
59335.8290	&	0.0002	&	16350.5	&	-0.0200	&	59349.1724	&	0.0002	&	16385	&	-0.0205	&	59363.0962	&	0.0002	&	16421	&	-0.0206	\\
59336.0221	&	0.0001	&	16351	&	-0.0204	&	59349.3668	&	0.0002	&	16385.5	&	-0.0195	&	59363.2910	&	0.0002	&	16421.5	&	-0.0192	\\
59336.2158	&	0.0001	&	16351.5	&	-0.0201	&	59349.5591	&	0.0002	&	16386	&	-0.0206	&	59363.4830	&	0.0002	&	16422	&	-0.0206	\\
59336.4089	&	0.0002	&	16352	&	-0.0204	&	59349.7535	&	0.0002	&	16386.5	&	-0.0196	&	59363.6778	&	0.0002	&	16422.5	&	-0.0192	\\
59336.6026	&	0.0002	&	16352.5	&	-0.0200	&	59349.9459	&	0.0001	&	16387	&	-0.0206	&	59363.8698	&	0.0002	&	16423	&	-0.0206	\\
59336.7957	&	0.0002	&	16353	&	-0.0204	&	59350.1403	&	0.0002	&	16387.5	&	-0.0196	&	59364.0647	&	0.0002	&	16423.5	&	-0.0191	\\
59336.9895	&	0.0002	&	16353.5	&	-0.0200	&	59350.3327	&	0.0002	&	16388	&	-0.0205	&	59364.2566	&	0.0002	&	16424	&	-0.0206	\\
59337.1824	&	0.0002	&	16354	&	-0.0204	&	59350.5271	&	0.0002	&	16388.5	&	-0.0195	&	59364.4515	&	0.0002	&	16424.5	&	-0.0191	\\
59337.3764	&	0.0002	&	16354.5	&	-0.0198	&	59350.7195	&	0.0002	&	16389	&	-0.0205	&	59364.6433	&	0.0002	&	16425	&	-0.0207	\\
59337.5691	&	0.0002	&	16355	&	-0.0205	&	59350.9138	&	0.0002	&	16389.5	&	-0.0196	&	59364.8382	&	0.0002	&	16425.5	&	-0.0192	\\
59337.7632	&	0.0002	&	16355.5	&	-0.0198	&	59351.1063	&	0.0002	&	16390	&	-0.0205	&	59365.0301	&	0.0002	&	16426	&	-0.0207	\\
59337.9559	&	0.0002	&	16356	&	-0.0205	&	59351.3006	&	0.0002	&	16390.5	&	-0.0196	&	59365.2249	&	0.0002	&	16426.5	&	-0.0192	\\
59338.1502	&	0.0002	&	16356.5	&	-0.0196	&	59351.4931	&	0.0002	&	16391	&	-0.0205	&	59365.4169	&	0.0002	&	16427	&	-0.0207	\\
59338.3425	&	0.0002	&	16357	&	-0.0206	&	59351.6874	&	0.0002	&	16391.5	&	-0.0196	&	59365.6117	&	0.0002	&	16427.5	&	-0.0193	\\
59338.5370	&	0.0002	&	16357.5	&	-0.0196	&	59351.8799	&	0.0002	&	16392	&	-0.0204	&	59365.8036	&	0.0002	&	16428	&	-0.0207	\\
59338.7293	&	0.0002	&	16358	&	-0.0206	&	59352.0742	&	0.0002	&	16392.5	&	-0.0195	&	59365.9985	&	0.0002	&	16428.5	&	-0.0192	\\
59338.9238	&	0.0002	&	16358.5	&	-0.0195	&	59352.2668	&	0.0002	&	16393	&	-0.0203	&	59366.1903	&	0.0002	&	16429	&	-0.0208	\\
59339.1161	&	0.0002	&	16359	&	-0.0206	&	59352.4609	&	0.0001	&	16393.5	&	-0.0196	&	59366.3853	&	0.0002	&	16429.5	&	-0.0192	\\
59339.3106	&	0.0002	&	16359.5	&	-0.0194	&	59352.6535	&	0.0001	&	16394	&	-0.0204	&	59366.5771	&	0.0002	&	16430	&	-0.0208	\\
59339.5029	&	0.0002	&	16360	&	-0.0205	&	59352.8477	&	0.0002	&	16394.5	&	-0.0196	&	59366.7720	&	0.0002	&	16430.5	&	-0.0192	\\
59339.6974	&	0.0002	&	16360.5	&	-0.0195	&	59353.0404	&	0.0001	&	16395	&	-0.0203	&	59366.9638	&	0.0002	&	16431	&	-0.0208	\\
59339.8898	&	0.0001	&	16361	&	-0.0205	&	59353.2346	&	0.0002	&	16395.5	&	-0.0195	&	59367.1590	&	0.0002	&	16431.5	&	-0.0191	\\
59340.0840	&	0.0002	&	16361.5	&	-0.0196	&	59353.4272	&	0.0002	&	16396	&	-0.0202	&	59367.3504	&	0.0002	&	16432	&	-0.0210	\\
59340.2767	&	0.0002	&	16362	&	-0.0204	&	59353.6213	&	0.0002	&	16396.5	&	-0.0195	&	59367.5458	&	0.0002	&	16432.5	&	-0.0190	\\
59340.4707	&	0.0002	&	16362.5	&	-0.0197	&	59353.8142	&	0.0002	&	16397	&	-0.0200	&	59367.7373	&	0.0002	&	16433	&	-0.0209	\\
59340.6636	&	0.0002	&	16363	&	-0.0202	&	59354.0081	&	0.0002	&	16397.5	&	-0.0195	&	59367.9325	&	0.0002	&	16433.5	&	-0.0191	\\
59340.8575	&	0.0002	&	16363.5	&	-0.0197	&	59354.2007	&	0.0001	&	16398	&	-0.0202	&	59368.1240	&	0.0002	&	16434	&	-0.0210	\\
59341.0503	&	0.0001	&	16364	&	-0.0203	&	59354.3949	&	0.0002	&	16398.5	&	-0.0195	&	59368.5108	&	0.0002	&	16435	&	-0.0210	\\
59341.2443	&	0.0002	&	16364.5	&	-0.0197	&	59354.5876	&	0.0002	&	16399	&	-0.0202	&	59368.7061	&	0.0002	&	16435.5	&	-0.0190	\\
59341.4370	&	0.0002	&	16365	&	-0.0203	&	59354.7816	&	0.0002	&	16399.5	&	-0.0195	&	59368.8976	&	0.0003	&	16436	&	-0.0209	\\
59341.6310	&	0.0002	&	16365.5	&	-0.0197	&	59354.9744	&	0.0001	&	16400	&	-0.0202	&	59369.0929	&	0.0002	&	16436.5	&	-0.0190	\\
59341.8238	&	0.0002	&	16366	&	-0.0204	&	59355.1683	&	0.0002	&	16400.5	&	-0.0196	&	59369.2844	&	0.0002	&	16437	&	-0.0209	\\
59342.0177	&	0.0002	&	16366.5	&	-0.0198	&	59355.3611	&	0.0001	&	16401	&	-0.0202	&	59369.4797	&	0.0002	&	16437.5	&	-0.0190	\\
59342.2105	&	0.0002	&	16367	&	-0.0204	&	59355.5551	&	0.0002	&	16401.5	&	-0.0196	&	59369.6712	&	0.0002	&	16438	&	-0.0209	\\
59342.4045	&	0.0002	&	16367.5	&	-0.0198	&	59355.7480	&	0.0002	&	16402	&	-0.0202	&	59369.8666	&	0.0002	&	16438.5	&	-0.0189	\\
59342.5972	&	0.0002	&	16368	&	-0.0205	&	59355.9419	&	0.0002	&	16402.5	&	-0.0196	&	59370.0580	&	0.0003	&	16439	&	-0.0209	\\
59342.7912	&	0.0002	&	16368.5	&	-0.0199	&	59356.1347	&	0.0002	&	16403	&	-0.0202	&	59370.2534	&	0.0002	&	16439.5	&	-0.0189	\\
59342.9840	&	0.0002	&	16369	&	-0.0205	&	59356.3286	&	0.0002	&	16403.5	&	-0.0197	&	59370.4448	&	0.0002	&	16440	&	-0.0208	\\
59343.1780	&	0.0002	&	16369.5	&	-0.0198	&	59356.5215	&	0.0001	&	16404	&	-0.0202	&	59370.6401	&	0.0002	&	16440.5	&	-0.0189	\\
59343.3707	&	0.0002	&	16370	&	-0.0205	&	59356.7154	&	0.0002	&	16404.5	&	-0.0197	&	59370.8317	&	0.0002	&	16441	&	-0.0207	\\
59343.5648	&	0.0001	&	16370.5	&	-0.0198	&	59356.9082	&	0.0002	&	16405	&	-0.0202	&	59371.0270	&	0.0002	&	16441.5	&	-0.0188	\\
59343.7574	&	0.0001	&	16371	&	-0.0206	&	59357.1022	&	0.0002	&	16405.5	&	-0.0197	&	59371.2185	&	0.0002	&	16442	&	-0.0207	\\
59343.9517	&	0.0002	&	16371.5	&	-0.0198	&	59357.2950	&	0.0002	&	16406	&	-0.0202	&	59371.4137	&	0.0002	&	16442.5	&	-0.0189	\\
59344.1442	&	0.0002	&	16372	&	-0.0206	&	59357.4889	&	0.0002	&	16406.5	&	-0.0197	&	59371.6053	&	0.0002	&	16443	&	-0.0206	\\
59344.3384	&	0.0002	&	16372.5	&	-0.0197	&	59357.6818	&	0.0001	&	16407	&	-0.0202	&	59371.8004	&	0.0002	&	16443.5	&	-0.0190	\\
59344.5309	&	0.0002	&	16373	&	-0.0206	&	59357.8757	&	0.0002	&	16407.5	&	-0.0197	&	59371.9921	&	0.0002	&	16444	&	-0.0207	\\
59344.7252	&	0.0002	&	16373.5	&	-0.0197	&	59358.0686	&	0.0001	&	16408	&	-0.0202	&	59372.1872	&	0.0003	&	16444.5	&	-0.0189	\\
59344.9178	&	0.0002	&	16374	&	-0.0206	&	59358.2625	&	0.0002	&	16408.5	&	-0.0197	&	59372.3788	&	0.0002	&	16445	&	-0.0207	\\
59345.1119	&	0.0002	&	16374.5	&	-0.0198	&	59358.4552	&	0.0002	&	16409	&	-0.0203	&	59372.5739	&	0.0002	&	16445.5	&	-0.0190	\\
59345.3046	&	0.0002	&	16375	&	-0.0206	&	59358.6493	&	0.0002	&	16409.5	&	-0.0197	&	59372.7656	&	0.0002	&	16446	&	-0.0207	\\
59345.4987	&	0.0002	&	16375.5	&	-0.0198	&	59358.8420	&	0.0001	&	16410	&	-0.0203	&	59372.9607	&	0.0002	&	16446.5	&	-0.0190	\\
59345.6914	&	0.0002	&	16376	&	-0.0205	&	59359.2287	&	0.0001	&	16411	&	-0.0204	&	59373.1524	&	0.0002	&	16447	&	-0.0206	\\
59345.8855	&	0.0002	&	16376.5	&	-0.0198	&	59359.4229	&	0.0002	&	16411.5	&	-0.0196	&	59373.5392	&	0.0002	&	16448	&	-0.0207	\\
59346.0782	&	0.0002	&	16377	&	-0.0204	&	59359.6155	&	0.0002	&	16412	&	-0.0204	&	59373.7342	&	0.0002	&	16448.5	&	-0.0190	\\
\hline
\end{tabular}
\end{center}
\label{appendix-table1}
\end{table*}

\renewcommand{\thetable}{A\arabic{table}}
\setcounter{table}{0}
\begin{table*}
\renewcommand\arraystretch{0.77}
\caption{Continued.}
\centering
\begin{center}
\begin{tabular}{c c c c c c c c c c c c}
\hline
Min.(BJD) & Error & Epoch & O-C & Min. & Error & Epoch & O-C & Min.(BJD) & Error & Epoch & O-C\\
\hline
59373.9259	&	0.0002	&	16449	&	-0.0207	&	59388.8182	&	0.0002	&	16487.5	&	-0.0194	&	60081.3371	&	0.0002	&	18278	&	-0.0247	\\
59374.1210	&	0.0002	&	16449.5	&	-0.0190	&	59389.0098	&	0.0003	&	16488	&	-0.0211	&	60081.7238	&	0.0002	&	18279	&	-0.0248	\\
59374.3127	&	0.0002	&	16450	&	-0.0207	&	59389.2048	&	0.0002	&	16488.5	&	-0.0195	&	60081.9200	&	0.0001	&	18279.5	&	-0.0219	\\
59374.5077	&	0.0002	&	16450.5	&	-0.0191	&	59389.3966	&	0.0003	&	16489	&	-0.0211	&	60082.1106	&	0.0002	&	18280	&	-0.0247	\\
59374.6990	&	0.0003	&	16451	&	-0.0212	&	59389.5922	&	0.0002	&	16489.5	&	-0.0189	&	60082.3068	&	0.0001	&	18280.5	&	-0.0219	\\
59376.0547	&	0.0002	&	16454.5	&	-0.0192	&	60068.7693	&	0.0002	&	18245.5	&	-0.0222	&	60082.4974	&	0.0002	&	18281	&	-0.0247	\\
59376.2464	&	0.0002	&	16455	&	-0.0209	&	60068.9606	&	0.0002	&	18246	&	-0.0243	&	60082.8841	&	0.0002	&	18282	&	-0.0248	\\
59376.6333	&	0.0002	&	16456	&	-0.0208	&	60069.1565	&	0.0001	&	18246.5	&	-0.0218	&	60083.0804	&	0.0002	&	18282.5	&	-0.0218	\\
59376.8282	&	0.0002	&	16456.5	&	-0.0192	&	60069.3474	&	0.0002	&	18247	&	-0.0243	&	60083.2708	&	0.0002	&	18283	&	-0.0249	\\
59377.0200	&	0.0002	&	16457	&	-0.0208	&	60069.5432	&	0.0001	&	18247.5	&	-0.0218	&	60083.8540	&	0.0001	&	18284.5	&	-0.0218	\\
59377.4068	&	0.0002	&	16458	&	-0.0208	&	60069.7341	&	0.0002	&	18248	&	-0.0243	&	60084.0444	&	0.0002	&	18285	&	-0.0248	\\
59377.6019	&	0.0002	&	16458.5	&	-0.0191	&	60069.9300	&	0.0001	&	18248.5	&	-0.0219	&	60084.2407	&	0.0001	&	18285.5	&	-0.0218	\\
59377.7936	&	0.0002	&	16459	&	-0.0208	&	60070.1209	&	0.0002	&	18249	&	-0.0243	&	60084.4311	&	0.0002	&	18286	&	-0.0248	\\
59377.9886	&	0.0002	&	16459.5	&	-0.0192	&	60070.3167	&	0.0001	&	18249.5	&	-0.0219	&	60084.6275	&	0.0001	&	18286.5	&	-0.0219	\\
59378.1804	&	0.0002	&	16460	&	-0.0207	&	60070.5077	&	0.0002	&	18250	&	-0.0243	&	60084.8179	&	0.0002	&	18287	&	-0.0249	\\
59378.5672	&	0.0002	&	16461	&	-0.0208	&	60070.7035	&	0.0001	&	18250.5	&	-0.0219	&	60085.0143	&	0.0001	&	18287.5	&	-0.0218	\\
59378.7622	&	0.0002	&	16461.5	&	-0.0191	&	60070.8945	&	0.0002	&	18251	&	-0.0243	&	60085.2047	&	0.0002	&	18288	&	-0.0248	\\
59378.9539	&	0.0002	&	16462	&	-0.0208	&	60071.0902	&	0.0001	&	18251.5	&	-0.0219	&	60085.4012	&	0.0001	&	18288.5	&	-0.0217	\\
59379.1489	&	0.0002	&	16462.5	&	-0.0192	&	60071.2812	&	0.0002	&	18252	&	-0.0243	&	60085.5915	&	0.0002	&	18289	&	-0.0248	\\
59379.3408	&	0.0002	&	16463	&	-0.0207	&	60071.4769	&	0.0001	&	18252.5	&	-0.0220	&	60085.7880	&	0.0002	&	18289.5	&	-0.0217	\\
59379.5356	&	0.0002	&	16463.5	&	-0.0193	&	60071.6680	&	0.0002	&	18253	&	-0.0243	&	60085.9783	&	0.0002	&	18290	&	-0.0248	\\
59379.7274	&	0.0002	&	16464	&	-0.0209	&	60071.8637	&	0.0001	&	18253.5	&	-0.0220	&	60086.1747	&	0.0002	&	18290.5	&	-0.0217	\\
59379.9224	&	0.0002	&	16464.5	&	-0.0192	&	60072.0548	&	0.0002	&	18254	&	-0.0243	&	60086.3651	&	0.0002	&	18291	&	-0.0248	\\
59380.1141	&	0.0002	&	16465	&	-0.0210	&	60072.2505	&	0.0001	&	18254.5	&	-0.0220	&	60086.7518	&	0.0002	&	18292	&	-0.0248	\\
59380.3092	&	0.0002	&	16465.5	&	-0.0192	&	60072.4416	&	0.0002	&	18255	&	-0.0243	&	60086.9483	&	0.0001	&	18292.5	&	-0.0218	\\
59380.5008	&	0.0002	&	16466	&	-0.0210	&	60072.6373	&	0.0001	&	18255.5	&	-0.0220	&	60087.1387	&	0.0002	&	18293	&	-0.0247	\\
59380.6961	&	0.0002	&	16466.5	&	-0.0192	&	60072.8283	&	0.0002	&	18256	&	-0.0244	&	60087.3350	&	0.0002	&	18293.5	&	-0.0218	\\
59380.8875	&	0.0002	&	16467	&	-0.0211	&	60073.0241	&	0.0001	&	18256.5	&	-0.0219	&	60087.5254	&	0.0002	&	18294	&	-0.0248	\\
59381.0828	&	0.0003	&	16467.5	&	-0.0192	&	60073.2150	&	0.0002	&	18257	&	-0.0244	&	60087.7218	&	0.0001	&	18294.5	&	-0.0218	\\
59381.2742	&	0.0003	&	16468	&	-0.0211	&	60073.4108	&	0.0001	&	18257.5	&	-0.0221	&	60087.9121	&	0.0002	&	18295	&	-0.0248	\\
59381.4696	&	0.0002	&	16468.5	&	-0.0192	&	60073.6018	&	0.0002	&	18258	&	-0.0244	&	60088.1085	&	0.0001	&	18295.5	&	-0.0219	\\
59381.6611	&	0.0003	&	16469	&	-0.0211	&	60073.7976	&	0.0001	&	18258.5	&	-0.0220	&	60088.2988	&	0.0002	&	18296	&	-0.0249	\\
59381.8564	&	0.0002	&	16469.5	&	-0.0191	&	60073.9886	&	0.0002	&	18259	&	-0.0244	&	60088.6856	&	0.0002	&	18297	&	-0.0249	\\
59382.0478	&	0.0003	&	16470	&	-0.0212	&	60074.1843	&	0.0001	&	18259.5	&	-0.0221	&	60088.8820	&	0.0001	&	18297.5	&	-0.0219	\\
59382.2433	&	0.0002	&	16470.5	&	-0.0190	&	60074.3753	&	0.0002	&	18260	&	-0.0244	&	60089.0724	&	0.0002	&	18298	&	-0.0249	\\
59382.4344	&	0.0003	&	16471	&	-0.0213	&	60074.5710	&	0.0001	&	18260.5	&	-0.0221	&	60089.2687	&	0.0001	&	18298.5	&	-0.0219	\\
59382.6299	&	0.0003	&	16471.5	&	-0.0192	&	60074.7622	&	0.0002	&	18261	&	-0.0244	&	60089.4591	&	0.0002	&	18299	&	-0.0250	\\
59382.8218	&	0.0002	&	16472	&	-0.0207	&	60074.9577	&	0.0001	&	18261.5	&	-0.0222	&	60089.6556	&	0.0002	&	18299.5	&	-0.0219	\\
59383.0167	&	0.0002	&	16472.5	&	-0.0192	&	60075.1490	&	0.0002	&	18262	&	-0.0243	&	60089.8459	&	0.0002	&	18300	&	-0.0250	\\
59383.2081	&	0.0002	&	16473	&	-0.0212	&	60075.3445	&	0.0001	&	18262.5	&	-0.0222	&	60090.0422	&	0.0001	&	18300.5	&	-0.0220	\\
59383.5949	&	0.0003	&	16474	&	-0.0212	&	60075.5357	&	0.0002	&	18263	&	-0.0244	&	60090.2334	&	0.0001	&	18301	&	-0.0242	\\
59383.7903	&	0.0003	&	16474.5	&	-0.0191	&	60075.7313	&	0.0001	&	18263.5	&	-0.0222	&	60090.6195	&	0.0002	&	18302	&	-0.0249	\\
59383.9816	&	0.0003	&	16475	&	-0.0212	&	60075.9226	&	0.0001	&	18264	&	-0.0242	&	60091.0062	&	0.0002	&	18303	&	-0.0250	\\
59384.3684	&	0.0002	&	16476	&	-0.0212	&	60076.3091	&	0.0002	&	18265	&	-0.0245	&	60091.2026	&	0.0001	&	18303.5	&	-0.0220	\\
59384.5636	&	0.0002	&	16476.5	&	-0.0193	&	60076.5049	&	0.0001	&	18265.5	&	-0.0221	&	60091.3930	&	0.0002	&	18304	&	-0.0250	\\
59384.7552	&	0.0002	&	16477	&	-0.0212	&	60076.6959	&	0.0002	&	18266	&	-0.0245	&	60091.7797	&	0.0002	&	18305	&	-0.0250	\\
59384.9504	&	0.0002	&	16477.5	&	-0.0194	&	60077.0827	&	0.0002	&	18267	&	-0.0245	&	60091.9762	&	0.0002	&	18305.5	&	-0.0219	\\
59385.1420	&	0.0003	&	16478	&	-0.0211	&	60077.2784	&	0.0001	&	18267.5	&	-0.0222	&	60092.1665	&	0.0002	&	18306	&	-0.0250	\\
59385.5288	&	0.0003	&	16479	&	-0.0212	&	60077.4695	&	0.0002	&	18268	&	-0.0245	&	60092.3630	&	0.0002	&	18306.5	&	-0.0219	\\
59385.7238	&	0.0002	&	16479.5	&	-0.0195	&	60077.6653	&	0.0001	&	18268.5	&	-0.0221	&	60092.5533	&	0.0002	&	18307	&	-0.0250	\\
59385.9156	&	0.0002	&	16480	&	-0.0211	&	60077.8563	&	0.0002	&	18269	&	-0.0245	&	60092.7498	&	0.0002	&	18307.5	&	-0.0219	\\
59386.1106	&	0.0002	&	16480.5	&	-0.0195	&	60078.2431	&	0.0002	&	18270	&	-0.0245	&	60092.9400	&	0.0002	&	18308	&	-0.0251	\\
59386.3024	&	0.0003	&	16481	&	-0.0211	&	60078.6298	&	0.0002	&	18271	&	-0.0245	&	60093.3268	&	0.0002	&	18309	&	-0.0251	\\
59386.4976	&	0.0002	&	16481.5	&	-0.0193	&	60078.8257	&	0.0001	&	18271.5	&	-0.0221	&	60093.5234	&	0.0002	&	18309.5	&	-0.0218	\\
59386.6892	&	0.0003	&	16482	&	-0.0211	&	60079.0166	&	0.0002	&	18272	&	-0.0245	&	60093.7136	&	0.0002	&	18310	&	-0.0251	\\
59387.0760	&	0.0003	&	16483	&	-0.0210	&	60079.2125	&	0.0001	&	18272.5	&	-0.0220	&	60093.9103	&	0.0002	&	18310.5	&	-0.0217	\\
59387.2712	&	0.0002	&	16483.5	&	-0.0192	&	60079.4033	&	0.0002	&	18273	&	-0.0246	&	60094.1004	&	0.0002	&	18311	&	-0.0250	\\
59387.4627	&	0.0003	&	16484	&	-0.0211	&	60079.7900	&	0.0002	&	18274	&	-0.0246	&	60094.4873	&	0.0002	&	18312	&	-0.0249	\\
59387.6579	&	0.0003	&	16484.5	&	-0.0194	&	60079.9861	&	0.0001	&	18274.5	&	-0.0220	&	60094.6839	&	0.0002	&	18312.5	&	-0.0217	\\
59387.8496	&	0.0003	&	16485	&	-0.0210	&	60080.1768	&	0.0002	&	18275	&	-0.0246	&	60094.8741	&	0.0002	&	18313	&	-0.0248	\\
59388.0446	&	0.0002	&	16485.5	&	-0.0194	&	60080.5635	&	0.0002	&	18276	&	-0.0247	&	60095.2609	&	0.0002	&	18314	&	-0.0249	\\
59388.2362	&	0.0003	&	16486	&	-0.0211	&	60080.7596	&	0.0001	&	18276.5	&	-0.0220	&	60095.4573	&	0.0002	&	18314.5	&	-0.0218	\\
59388.4314	&	0.0002	&	16486.5	&	-0.0194	&	60080.9503	&	0.0002	&	18277	&	-0.0247	&	60095.6477	&	0.0002	&	18315	&	-0.0248	\\
59388.6231	&	0.0002	&	16487	&	-0.0210	&	60081.1464	&	0.0001	&	18277.5	&	-0.0219	&	60096.0344	&	0.0002	&	18316	&	-0.0249	\\
\hline
\end{tabular}
\end{center}
\label{appendix-table2}
\end{table*}

\clearpage

\bibliography{REFS}{}
\bibliographystyle{aasjournal}

\end{document}